\documentclass{elsart}

\usepackage{lineno}

\usepackage{epsfig}

\usepackage{amssymb}

\begin{document}

\begin{frontmatter}



\title{Monte Carlo simulations of air showers in atmospheric electric
fields}
\author[nijmegen]{S.~Buitink\corauthref{cor1}}
\ead{s.buitink@astro.ru.nl},
\author[fzk]{T.~Huege},
\author[nijmegen]{H.~Falcke},
\author[fzk]{D.~Heck},
\author[nijmegen]{J.~Kuijpers}
\corauth[cor1]{Corresponding author.}
\address[nijmegen]{Department of Astrophysics/IMAPP, Radboud University Nijmegen, P.O.~Box 9010, 
6500 GL Nijmegen, The Netherlands}
\address[fzk]{Institut f\"ur Kernphysik, Forschungszentrum Karlsruhe,
76021 Karlsruhe, Germany}





\begin{abstract}
The development of cosmic ray air showers can be influenced by atmospheric
electric fields. Under fair weather conditions these fields are small, but the
strong fields inside thunderstorms can have a significant effect on the
electromagnetic component of a shower. Understanding this effect is particularly
important for radio detection of air showers, since the radio emission is
produced by the shower electrons and positrons.
We perform Monte Carlo simulations to calculate the effects of different
electric field configurations on the shower development.
We find that the electric field becomes important for values of the order of 1 kV/cm. Not only can the energy distribution of electrons and positrons 
change significantly for such field strengths, it is also possible that runaway electron breakdown 
occurs at high altitudes, which is an important effect in lightning initiation. 
\end{abstract}


\end{frontmatter}

\section{Introduction}
\label{sec:intro}

The effect of atmospheric electric fields on the development of extensive air showers from high energy cosmic rays has not received much
attention in the past. 
Because of the large energies of shower particles, the electric fields present in the atmosphere are generally much too small to
alter the particle energies significantly. The largest fields are of the order of 1 kV/cm and only occur in
thunderstorms \cite{book}. In such fields the hadronic and muonic part of the shower are hardly affected, although a muon deficit due to increased
decay rate has been reported by Alexeenko et al.~\cite{A02}. The effects on the electromagnetic shower are much larger, but they are, 
as we will show in this work, local in the sense that the amount and energy distribution of
electromagnetic particles quickly adapts to the local background field.

Below thunderstorms the electric field decreases, so particle detector arrays will 
in general not be strongly sensitive to the influence of electric fields. Mountain top experiments, however, can be very close to, or even inside 
thunderstorms. An increase in the air shower detection rate and the single particle count has been reported by high altitude experiment such as
EAS-TOP \cite{eastop} and the Carpet air shower array \cite{A02}.  

Effects of atmospheric electric fields are generally not included in air shower simulation codes like CORSIKA \cite{CORSIKA}. There
are, however, two topics for which we demonstrate in this paper that it is important to take electric fields into consideration.

The first is the consequence of different shower developments in strong electric field regions on the radio emission of the shower. With
experiments such as LOPES \cite{F05} and Codalema \cite{codalema}, the technique of radio detection of air showers has become mature 
in the last few years. Radio antennas, with their low costs
and high duty cycles, have been demonstrated to be an attractive addition to large air shower arrays like the Pierre Auger Observatory
\cite{PAO}. Already in the 1970s, experimental results by Madolesi et al.~\cite{M74} showed excessively large radio pulses during
thunderstorms. Together with a large range in measured radio intensity and the inability to filter out radio interference, the 
unknown effect of electrical conditions in the atmosphere was a reason to abandon the radio experiments \cite{B77} until the development of
digital radio arrays like LOFAR \cite{lofar} revived the interest. A study of the effect of weather conditions on the radio pulse strength with LOPES data confirmed the
amplification of radio pulses during thunderstorm conditions, but also showed that no effect is observed under other weather
conditions \cite{B07}. In another study \cite{N08} it was shown that during thunderstorms the air shower arrival 
direction reconstructed with the LOPES radio antennas can be a few degrees off with respect to the direction reconstructed by
the KASCADE particle detector array.
  
It is now known that the radio emission of air showers can be described in terms of coherent synchrotron radiation of shower electrons and positrons that follow
curved trajectories in the Earth's magnetic field as first described in Falcke et al.~\cite{FG03} and in more detail in Huege et al.~\cite{HF03}. 
Alternatively, a macroscopic picture can be constructed in
which the shower charges, that are separated in the magnetic field, support a transverse current producing
radiation as proposed by Kahn \& Lerche \cite{KL66} and more recently by Scholten et al.~\cite{Sch}. Although there are subtle differences between 
these descriptions, it is clear that an alteration in the electron and positron distribution of the 
shower at some altitude, can influence the power of the radio pulse in both models. In this paper we limit ourselves to a study of electric field effects on
these distributions, not on the radio pulse itself, which will be the subject of a forthcoming paper.

The second issue related to the electric field is the suggestion that air showers of sufficient energy can 
start an avalanche of runaway electrons in thunderstorm electric fields. Ionization electrons that are produced in collisions of shower
particles with air molecules are accelerated in the thunderstorm electric field and can, under the right conditions, gain enough energy to
ionize further molecules, an effect described by Gurevich et al.~\cite{G92}. In thunderstorm research the field strength that can support
such avalanches is known as the threshold field, described in Marshall et al.~\cite{MMR95}. In their work, the authors present 
thunderstorm measurements which show that lightning
often occurs when the thunderstorm field exceeds the breakeven field, suggesting that runaway electron breakdown plays a role
in lightning initiation. By providing seed electrons for avalanches, air showers from cosmic rays may play an important role in thunderstorm dynamics. Simulations by Dwyer \cite{Dw03} have shown that the threshold field strength for an avalanche to develop is slightly higher than the breakeven field, taking into account the effects of elastic scattering.

Inside thunderstorms there have been measurements of X-ray bursts \cite{MP85} and gamma bursts, both from Earth \cite{G04} and space \cite{F94}.
This emission can be explained in terms of bremsstrahlung emitted by the runaway breakdown electrons \cite{GM99, Dw08a}.

In this work, we simulate the effect of electric fields on the development of air showers with CORSIKA. In Sec.~\ref{sec:sim} we describe
the setup of our simulation and the modification that has been made for CORSIKA. In Sec.~\ref{sec:effects} we derive some simple analytic
estimates and limits to compare with the results. Simulation results are presented in Sec.~\ref{sec:res} and we conclude with a discussion of
the simulation limitations and the consequences for realistic field configurations.

\section{Simulation Setup}
\label{sec:sim}

CORSIKA \cite{CORSIKA} is a Monte Carlo code that simulates the development of air showers by tracing the individual particles and their interactions 
based on several sophisticated interaction models. In our
simulations we use the high-energy hadronic interaction model QGSJET-II \cite{QGSJET} and for low-energy hadronic interactions we use UrQMD 1.3cr
\cite{URQMD}. 

Electromagnetic interactions are simulated by the standard
CORSIKA (version 6.720) routines to treat electromagnetic particles. These routines
are taylor-made versions of the EGS4-code \cite{EGS4} adapted to the barometric
atmosphere with a density decreasing exponentially with increasing
altitude. All possible interactions are considered and a proper
treatment of ionization energy loss and multiple scattering is performed.
By including some suitable extra statements into the transport routine
ELECTR for e$^{+/-}$ particles the effects of an external electrical
field are taken into account which causes an acceleration (energy
gain rsp. loss) for particles moving parallel to the field and
a deflection for those moving perpendicular to the field. A suitable
limitation of the transport step length guarantees small changes of
the particle movements to neglect higher order effects on the particle
traces. By these means the energy gain/loss in the electrical field
and the ionization energy loss can be treated independently for
each transport step.
In our simulations we use the ``thinning'' option with thinning at $10^{-7}$ level and optimized weight limitation \cite{kobal} to keep the computing
times in a tolerable level. The energy cutoff below which electrons and positrons are discarded from the simulation is
0.5~MeV. Because in an electric field, particles below this threshold may be accelerated
to higher energies, this introduces a limitation to the simulations, which is discussed in Sec.~\ref{sec:dis}.  

Simulations were done on five proton showers: vertical showers of $10^{16}$~eV, $10^{17}$~eV and respectively, $3\cdot 10^{17}$~eV and two inclined showers of
$10^{16}$~eV, having zenith angles of 30 and 60 degrees. All showers had a proton as primary particle. 
We use homogeneous ambient electric fields of strengths up to 1 kV/cm. In reality, fields of such strengths will only occur in small parts of a
thunderstorm, but keeping the field constant in the simulations allows us to understand the electric field effect better. 
In Sec.~\ref{sec:dis} we will discuss the implications for realistic field configurations. 

When simulating showers with the same primary particle but different random seeds fluctuations will occur from shower to shower. Most importantly the altitude of the first
interaction varies, but also the 
location of the shower maximum, for example, is dependent on number of particles that are produced in the first interaction, and the energy
distribution of these particles. We use CONEX \cite{CONEX} to make 100 shower simulations for each of the five above-mentioned configurations. 
From these 100 simulations we select a shower with a large number of secondary particles in the first interaction and a fairly typical longitudinal shower profile. CONEX produces a
file that lists all secondary particles after the first interaction and their momenta, which can be used as an input stack for CORSIKA using the STACKIN option. With a
CONEX stack of particles created at the first interaction instead of one primary particle as input, different random seeds will produce much smaller variations. For each shower
configuration we have selected a CONEX input stack and used this to produce ten showers with different random seeds. In the
following plots of shower evolution we plot the mean value of these ten showers and one sigma error bars.
Because the fluctuations between simulations are very small with this approach, we are more senstitive to changes that are introduced by the
background electric field. 

We use the COAST interface code for CORSIKA \cite{COAST} to get information on electron and positron distributions at 50 layers evenly
distributed in atmospheric depth along the shower axis. At each
layer two three-dimensional histograms are written out. One shows the distribution as a function of, respectively:
\begin{itemize}
\item particle energy,
\item distance of the particle to the shower axis and,
\item time lag of the particle with respect to a plane that travels along the shower axis with the speed of light in vacuum.
\end{itemize}
The second histogram shows the distirbution as a function of:
\begin{itemize}
\item particle energy,
\item angle between the particle momentum and shower axis and,
\item angle between the component of the particle momentum that is perpendicular to the shower axis and a vector pointing radially outwards
from the shower axis to the particle.
\end{itemize}
Both histograms are created separately for electrons and positrons.

\begin{figure}[htp]
\centerline{\epsfig{file=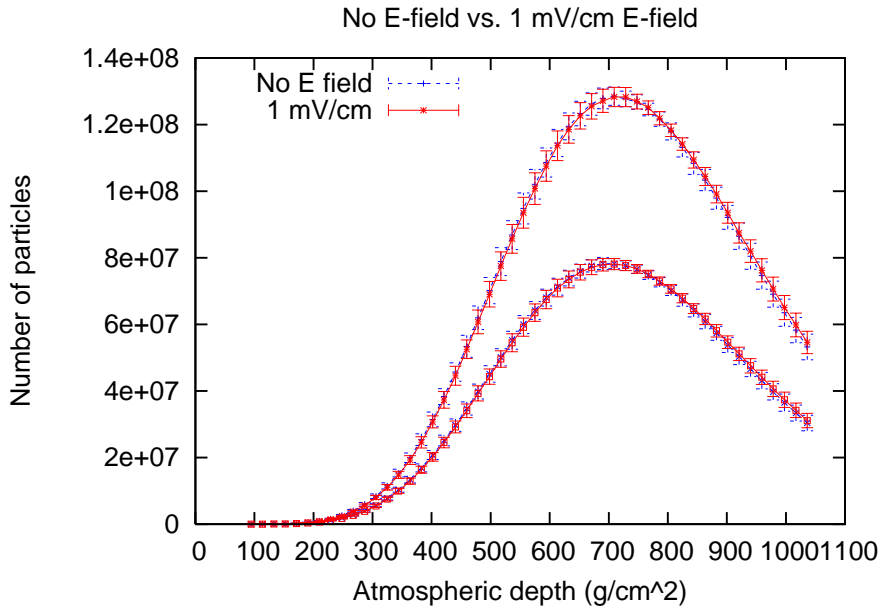 ,width=0.6\textwidth, angle=270}}
\caption{Shower evolution for a vertical shower of $3\cdot 10^{17}$~eV. The blue dashed lines represent simulations
with the original CORSIKA code in which no electric field is incorporated. The red solid lines 
represent simulations with a small electric field added of 1 mV/cm and are almost exactly on top of the blue dashed curves. The top lines show electrons and
the bottom lines show positrons.}
\label{NoE_vs_SmallE} 
\end{figure}

In Fig.~\ref{NoE_vs_SmallE}
the number of electrons and positrons is plotted as a function of atmospheric depth for a $3\cdot 10^{17}$~eV shower. Simulation results with the original code, not including an
electric field,  
are represented by the dashed blue line and the red solid line shows the results of the modified code, with a very small electric field added of 1~mV/cm. For small electric 
fields the original and 
modified CORSIKA codes converge to within the limit of random fluctuations. This is also observed for showers of different energy and zenith angle.  

\section{Electric field effects}
\label{sec:effects}

\begin{figure}[htp]
\centerline{\epsfig{file=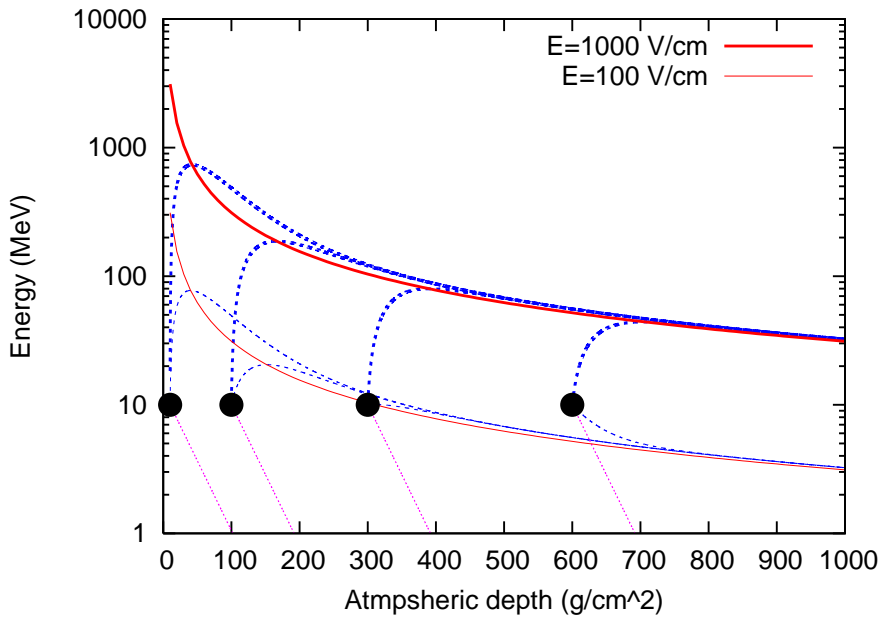 ,width=0.6\textwidth, angle=270}}
\caption{Red (solid) lines show equilibrium energy as a function of atmospheric depth for a field of 100 V/cm (thin) and 1000 V/cm (thick). Trajectories of
accelerated particles with smooth energy losses are plotted starting at the bullets. The blue (dashed) lines correspond to trajectories inside
electric fields of 100 V/cm (thin) and 1000 V/cm (thick) and the purple (dotted) line to trajectories in the absence of an electric field.}
\label{fig:equilibrium} 
\end{figure}

High energy electrons lose their energy primarily through the process of bremsstrahlung. The radiation length for
electrons in air is $X_0 \sim 36.7$~g/cm$^{-2}$ \cite{PDB}. When an electric field $E$ is present the change in energy $U$ per unit of atmospheric
depth $X$ is given by:
\begin{equation}
\mathrm{d}U=-\frac{U}{X_0}\mathrm{d}X - q E \mathrm{d}z = -\frac{U}{X_0} \mathrm{d}X+ \frac{q E z_0}{X} \mathrm{d}X,
\label{bremsloss}
\end{equation}
where $U$ is the particle energy and $q$ the particle charge. The vertical distance $\mathrm{d}z$ is related to the vertical atmospheric depth
as $\mathrm{d}z=-\rho^{-1} dX$. The air density $\rho$ is approximately given by \cite{MMR95}:
\begin{equation}
\rho(z)=1.208\cdot 10^{-3} \exp(-z/z_0) \textrm{g/cm$^3$},
\end{equation}
where $z_0\approx 8.4$~km is the scale height, so $\rho=X/z_0$, which is used in the last step of Eqn. \ref{bremsloss}. The particle reaches an equilibrium energy when
\begin{equation}
U(X)=\frac{q E z_0 X_0}{X}.
\label{equilibrium}
\end{equation}
Particles below this energy are accelerated, while for particles above this energy radiation losses dominate. Fig.~\ref{fig:equilibrium}
shows this equilibrium energy as well as trajectories (in energy) of particles that are created at a certain atmospheric depth with an
energy of 10 MeV and only lose energy through smooth bremsstrahlung losses as in Eqn.~\ref{bremsloss}. It can be seen that particles are 
accelerated until the point where they cross the equilibrium energy. Although this picture is a strong simplification of the realistic case where the
particles do not lose their energy smoothly over time but in discrete processes, we can use it to set two limits on the effect of an electric field
on the energy distribution of shower particles: the distribution will be
largely unaffected beyond the maximum energy that accelerated particles can reach (e.g.\ $\sim 200$~MeV for 1 kV/cm at X=100 g/cm$^2$, see
Fig.~\ref{fig:equilibrium}) and the effect
on the distribution will be most prominent below the equilibrium energy given by Eqn.~\ref{equilibrium}.

Below $\sim 1$ MeV, the energy loss due to collisions increases with decreasing energy. For a given electric field strength, only particles
above a certain energy can be accelerated. The breakeven field is defined as the field strength at which the energy loss due to collisions
is equal to the energy gain in the electric field for an electron of 1 MeV and is given by \cite{MMR95}: 
\begin{equation}
E_{be}(z)=1.67\cdot 10^6 \rho(z) \textrm{V/cm.}
\label{breakeven}
\end{equation}
Electrons with energies higher than 1 MeV will be accelerated in a breakeven field, while electrons with energy below 1 MeV will decelerate.
For fields exceeding the breakeven field, the mimimum electron energy required for acceleration will be lower.

In collisions with air molecules air shower particles produce many electrons with energies below 1 MeV. When the background electric field
exceeds the breakeven field a part of these electrons will be accelerated to energies at which they can create new ionization electrons,
resulting in a breakdown process. It has been found by Dwyer \cite{Dw03} that the threshold field strength for the development of a runaway breakdown process is:
\begin{equation}
E_{be}(z)=2.35\cdot 10^6 \rho(z) \textrm{V/cm.}
\label{threshold}
\end{equation}

A field of 1000 V/cm is equal to the threshold field at $z\approx 8.7$~km or $X\approx 350$~g/cm$^2$. Above this altitude we can encounter 
breakdown effects. When the breakdown is efficient it gives an exponential increase of the number of electrons. The breakdown electrons can be distinguished
from the electrons from pair creation in three ways: (a) their energies are generally lower, (b) as they are created at lower energies they will have a
larger time delay w.r.t. the shower front and (c) they are deflected strongly (or entirely) into the electric field direction.

At  $z\approx 5.8$~km or $X\approx 500$~g/cm$^2$, a field of 1000 V/cm is equal to the breakeven field. Below this altitude no breakdown effects are expected.

The complications of electron runaway breakdown are described in much detail in various publications, see for example Gurevich et al.~\cite{G92, G99, G06}. The 
simple features presented here will serve as a handle to interpret the simulation results.

\section{Simulation results}
\label{sec:res}
As follows from Eqns.~\ref{bremsloss} and \ref{equilibrium} a positive field points downwards and accelerates the positrons. The errors bars are 1
sigma variations of 10 simulations with the same CONEX input stack of particles after the first interaction. From these 10 showers we choose the one of which the longitudinal
profile is closest to the mean profile, when producing energy distribution plots.
\subsection{Vertical showers}
\label{sec:res.dev}
\begin{figure}[htp]
\centerline{\epsfig{file=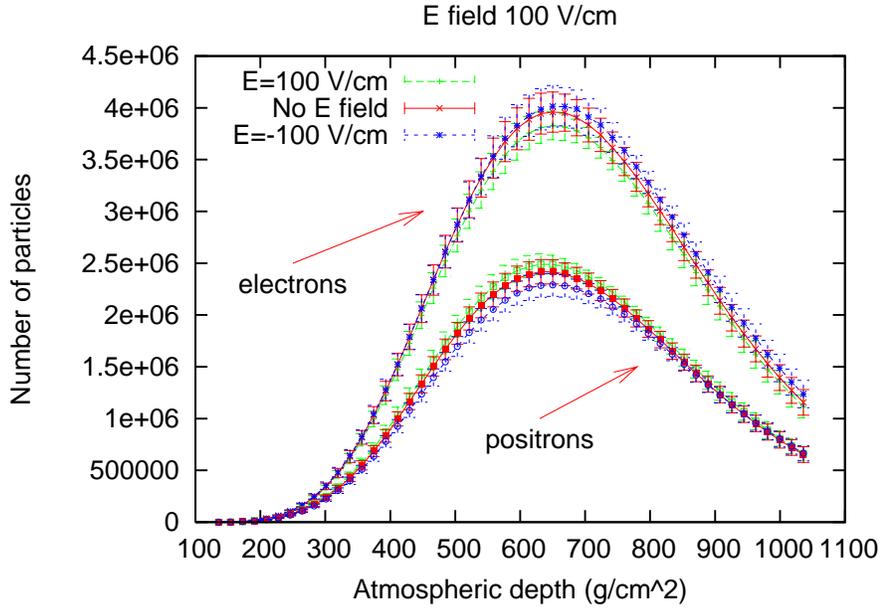 ,width=0.6\textwidth, angle=270}}
\caption{Number of electrons and positrons as a function of atmospheric depth for a 10$^{16}$ eV vertical shower for different
electric field strengths.}
\label{const_field_100} 
\end{figure}
Fig.~\ref{const_field_100} shows simulation results for a vertical $10^{16}$~eV air shower. The number of electrons and positrons is
plotted as a function of atmospheric depth. The red data points correspond to the absence of an electric field and the green
and blue points to fields of 100~V/cm and -100~V/cm respectively. The variations are within the 1 sigma error bars. 
\begin{figure}[htp]
\centerline{\epsfig{file=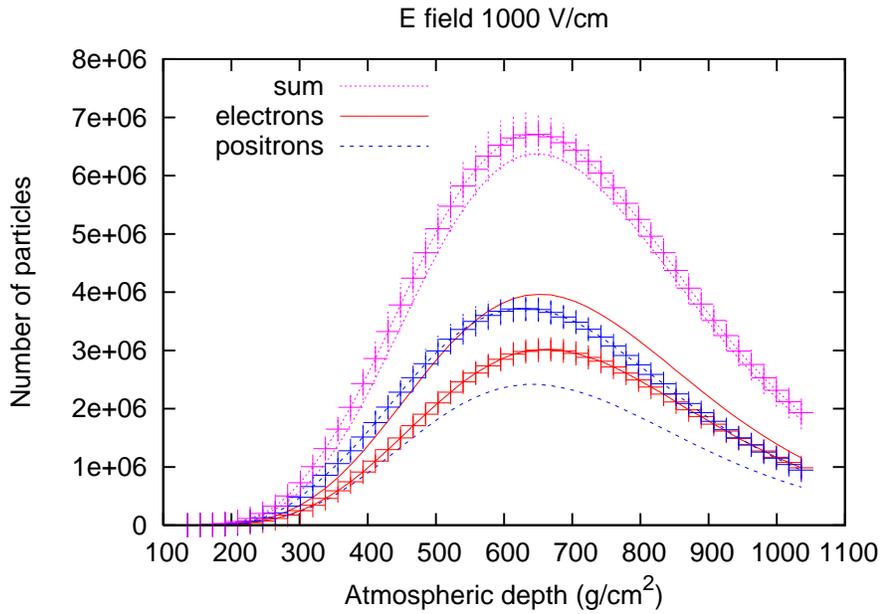 ,width=0.6\textwidth, angle=270}}
\caption{Number of electrons, positrons, and sum of both as a function of atmospheric depth for a 10$^{16}$ eV vertical shower inside an
electric field of 1000 V/cm. The lines without error bars represent the shower development in absence of a field.}
\label{const_field_1000} 
\end{figure}

\begin{figure}[htp]
\centerline{\epsfig{file=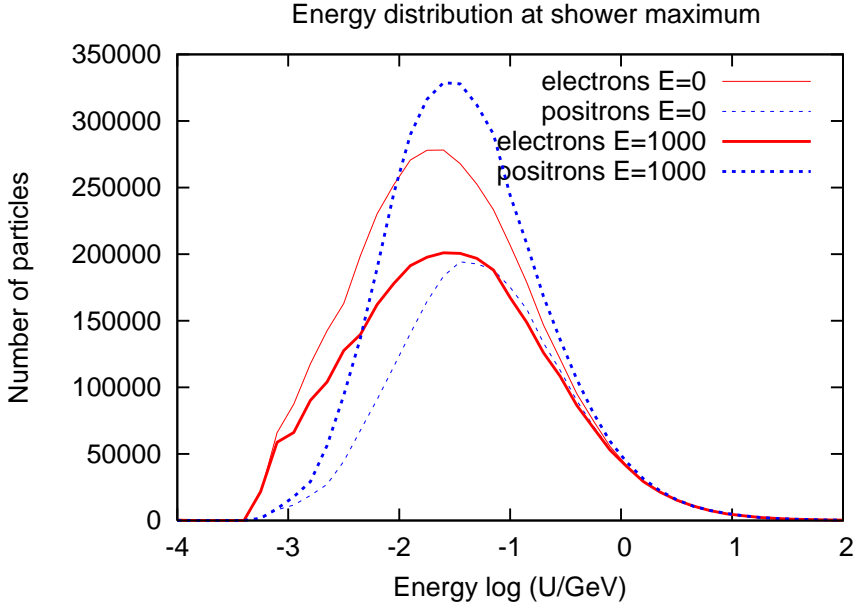 ,width=0.6\textwidth, angle=270}}
\caption{Energy distribution of a 10$^{16}$ eV vertical shower inside an
electric field of 1000 V/cm at the shower maximum.}
\label{enerdist1000} 
\end{figure}

When the field strength is increased by an order of magnitude the effect on the shower development becomes significant.
Fig.~\ref{const_field_1000} shows the number of electrons and positrons for a $10^{16}$~eV vertical shower in an
electric field of 1000 V/cm (accelerating the positrons). The lines without error bars represent the development in absence of an electric field.
When the field is switched on, the number of positrons outgrows the number of electrons, causing a positive charge
excess. Fig.~\ref{enerdist1000} shows the energy distribution of electrons and positrons at the shower maximum. The
distributions for the same shower in absence of an electric field is plotted for reference. The sum of electron and positrons
increases slightly but within the 1 $\sigma$ error region.
 
\begin{figure}[htp]
\centerline{\epsfig{file=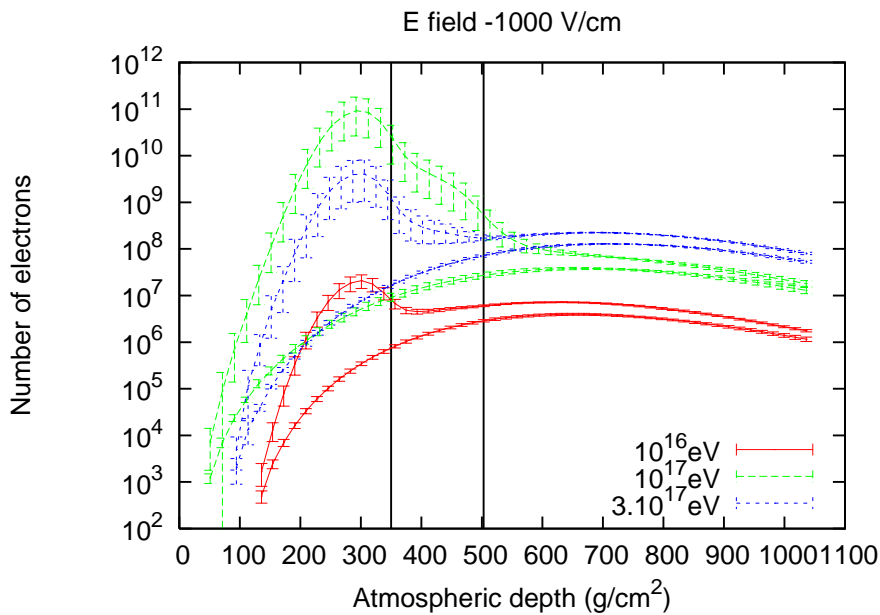 ,width=0.6\textwidth, angle=270}}
\caption{Number of electrons as a function of atmospheric depth for vertical showers of several energies. For each energy
the lower line represents the shower development in the absence of a field. The vertical solid line marks the altitude at which the electric field
equals the threshold field (left line) and breakeven field (right line).}
\label{const_field_-1000} 
\end{figure}

Fig.~\ref{const_field_-1000} shows the shower development in an electric field of -1000 V/cm (accelerating the electrons) for several shower energies. For each energy
the lower line represents the same shower in absence of a field. Above the altitude in which the electric field equals the threshold field (Eqn.\ \ref{threshold}) 
an explosive increase in the number of electrons can be seen (note we switched to logaritmic scale). High up in the atmosphere the
increase of electrons is nearly exponential. Interestingly, the largest electron content is reached by the shower
that had its primary interaction highest up in the atmosphere, not the shower with the highest primary energy. The latter does have the most electrons at 
lower altitudes, below the altitude at which the electric field equals the breakeven field, where the breakdown process has stopped and the electrons are injected by pion decay. High up in the atmosphere, the number of
electrons increases exponentially reaching a maximum at $X\approx 300$~g/cm$^2$, where the electric field is slightly stronger than the threshold field. The energy cutoff at 0.5 MeV in our simulation may be of influence to the location of the maximum. 
Note that the point of first interaction for the showers of different energies is random due to the way we selected our showers, and does not follow
the dependence of mean first interaction height on primary energy.

\begin{figure}[htp]
\centerline{\epsfig{file=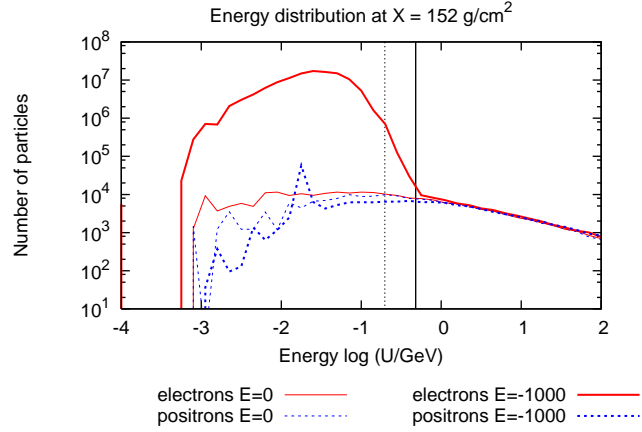 ,width=0.6\textwidth}}
\caption{Energy distribution of a 10$^{17}$ eV vertical shower inside an upward (negatively aligned)
electric field of 1000 V/cm and the same shower in absence of a field at X=152 g/cm$^2$. The vertical lines represent the maximum energy a particle
can have gained in the electric field (solid) and the equilibrium energy (dotted).}
\label{enerX152} 
\end{figure}

\begin{figure}[htp]
\centerline{\epsfig{file=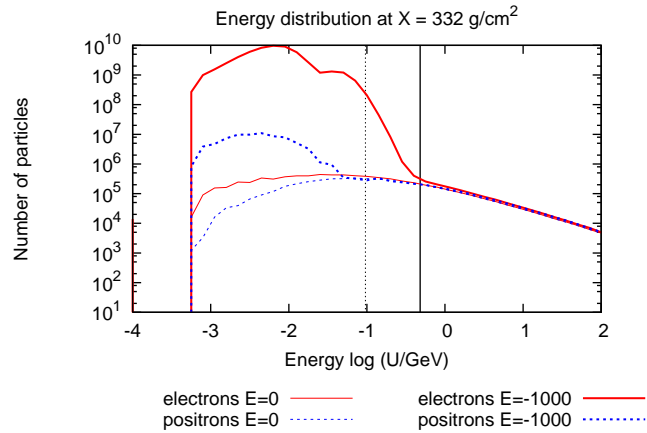 ,width=0.6\textwidth}}
\caption{Same as Fig.~\ref{enerX152} at X=332 g/cm$^2$.}
\label{enerX332} 
\end{figure}

\begin{figure}[htp]
\centerline{\epsfig{file=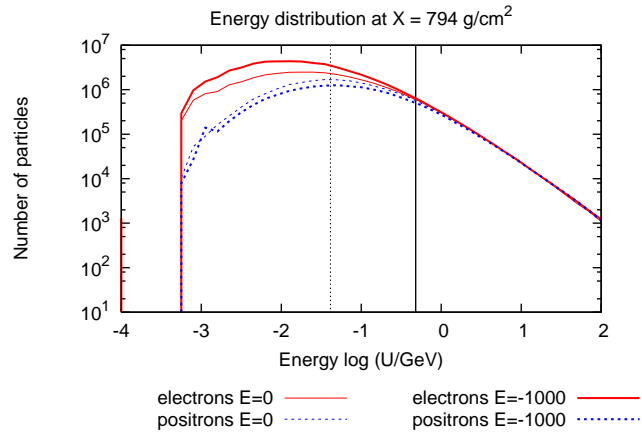 ,width=0.6\textwidth}}
\caption{Same as Fig.~\ref{enerX152} at X=794 g/cm$^2$.}
\label{enerX794} 
\end{figure}

Figs.~\ref{enerX152} through \ref{enerX794} show the energy distribution of electrons and positrons for the $10^{17}$~eV shower in Fig.~\ref{const_field_-1000}. 
The same shower in absence of a field is plotted for reference. Two vertical lines represent the limits derived in Sec.~\ref{sec:effects}: the main effect of the 
particle acceleration is expected to occur below the equilibrium energy (Eqn.~\ref{equilibrium}) and no significant change is
expected above the equilibrium energy at the altitude of first interaction (dotted and solid line). The plots represent the early breakdown phase (Fig.~\ref{enerX152}), 
the altitude at which the number of electrons reaches its maximum (Fig.~\ref{enerX332}) and the altitude at which the shower reaches its maximum in absence of a field (Fig.~\ref{enerX794}). At the point where the number of electrons reaches its maximum, the amount of positrons is also increased by a
factor 100. Apparently, the number of electrons in the breakdown process is so large that the pair creation from gamma emission of the 
breakdown electrons exceeds the pair creation from gamma emission of pion decay.

\begin{figure}[htp]
\centerline{\epsfig{file=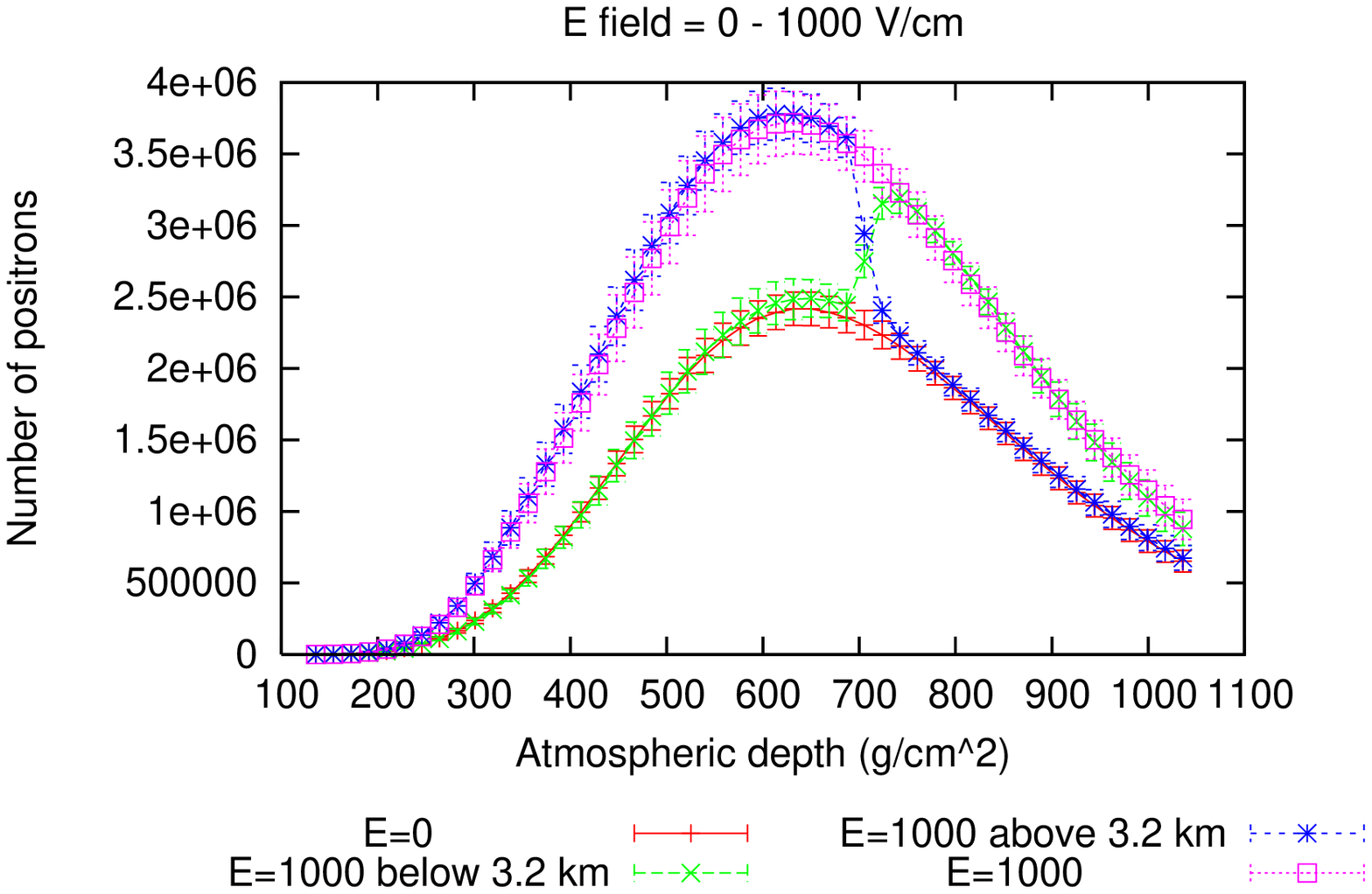 ,width=0.6\textwidth, angle=270}}
\caption{Positron number development of a 10$^{16}$~eV shower inside different electric field configurations.}
\label{localeffect} 
\end{figure}
To study the effect of a local electric field, two field configurations are simulated in which the electric field is switched on or off, respectively, at the altitude
of 3.2 km.
Fig.~\ref{localeffect} shows the number of positrons as a function of atmospheric depth for a 10$^{16}$~eV shower in absence of a field, 
inside a field of 1000 V/cm that is homogeneous over the whole atmosphere and inside two fields that have a discontinuous jump at 3.2 km
altitude. One of the latter has field strength 1000 V/cm above 3.2 km and no electric field below that altitude, the other vice versa.
It can be seen
that the
number of particles quickly adapts itself to the expected number for the local electric field. The reason is that the electromagnetic component is
continually refreshed by pair production due to pion decay, while the electric particles lose their energy quickly through radiation and ionization losses.    

\subsection{Inclined showers}
\label{sec:res.inc}
\begin{figure}[htp]
\centerline{\epsfig{file=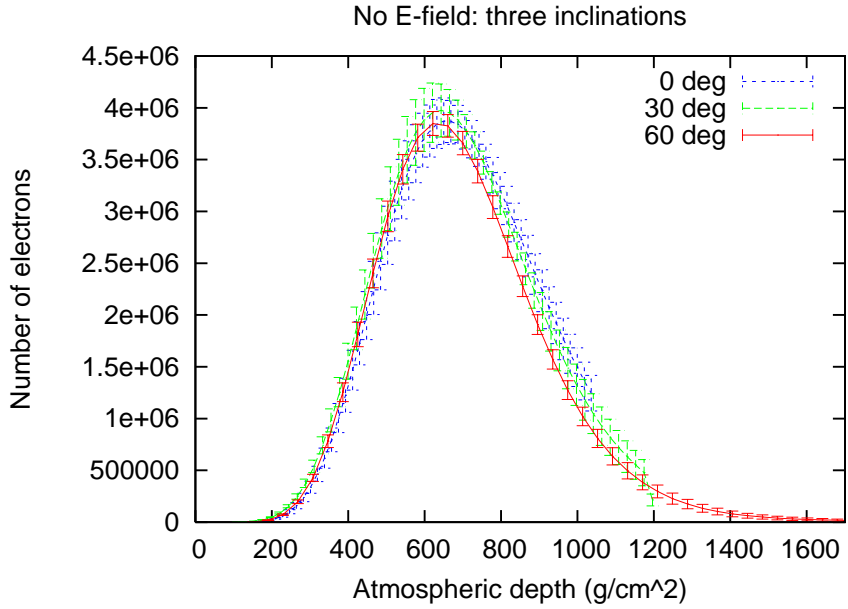 ,width=0.6\textwidth, angle=270}}
\caption{Shower development of $10^{16}$~eV showers of three different inclinations in absence of an electric field.} 
\label{nofield3i}
\end{figure}

\begin{figure}[htp]
\centerline{\epsfig{file=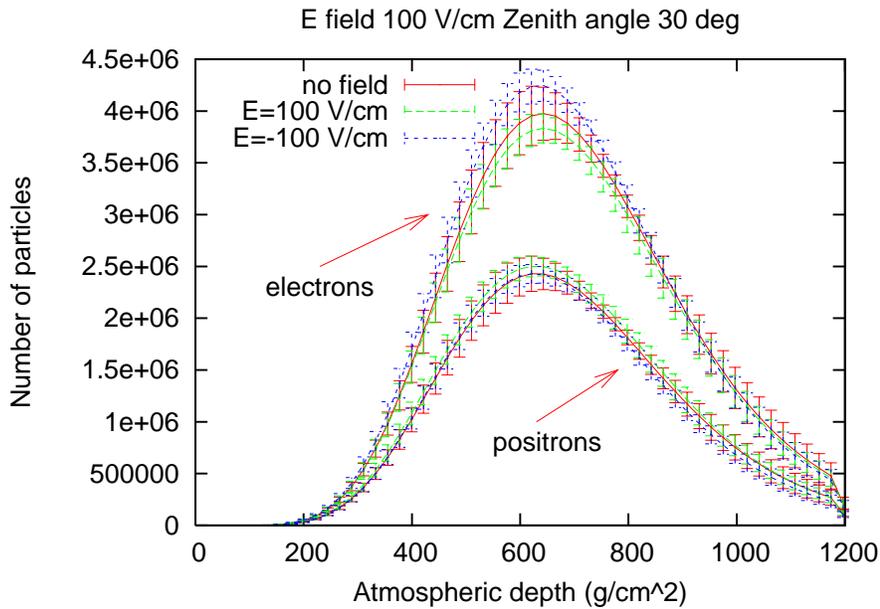 ,width=0.6\textwidth, angle=270}}
\caption{Shower development of an inclined $10^{16}$~eV shower with 30 degrees zenith angle in different electric fields.} 
\label{smallfield30}
\end{figure}

\begin{figure}[htp]
\centerline{\epsfig{file=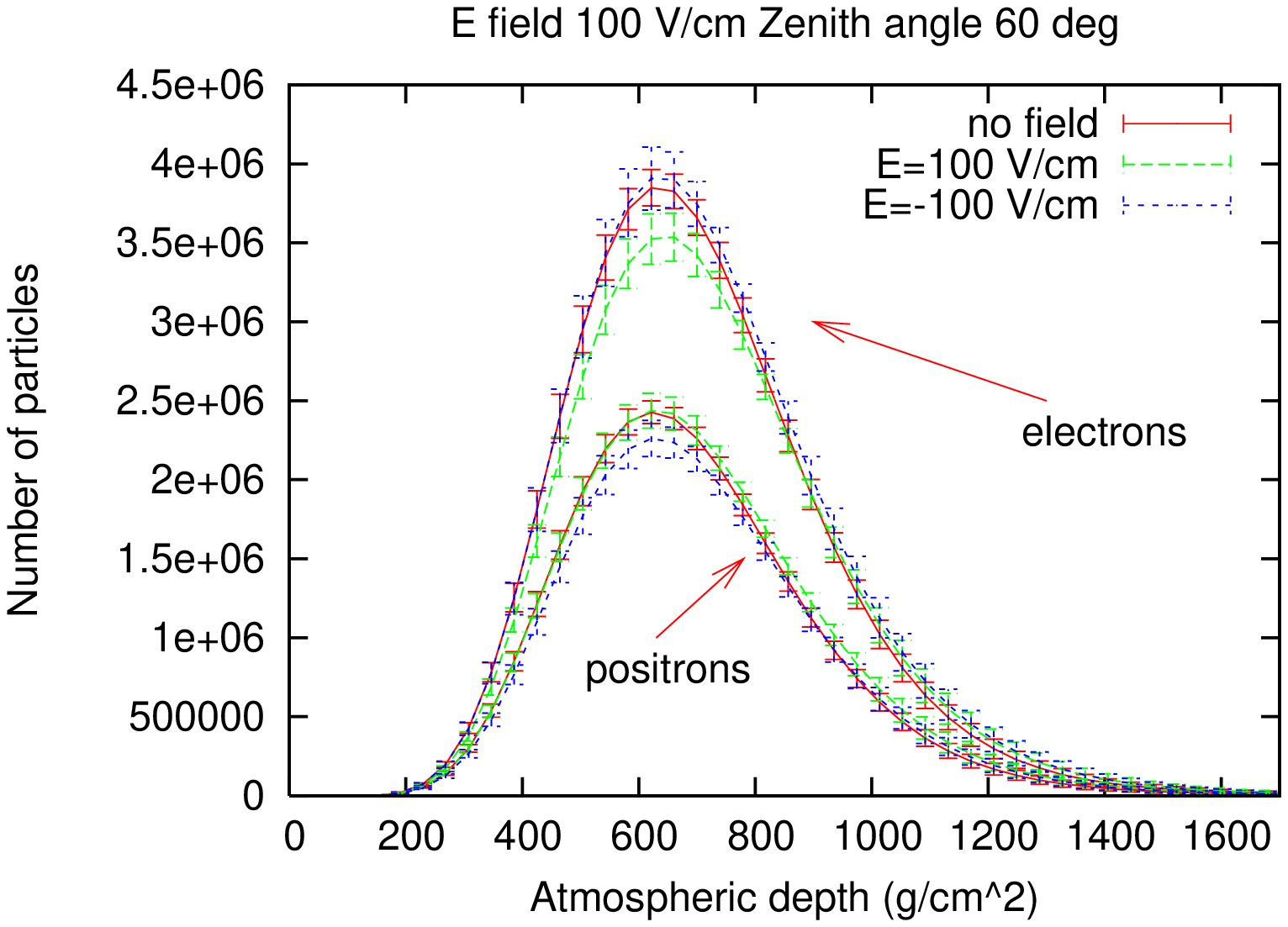 ,width=0.6\textwidth, angle=270}}
\caption{Shower development of an inclined $10^{16}$~eV shower with 60 degrees zenith angle in different electric fields.} 
\label{smallfield60}
\end{figure}

For inclined showers the longitudinal development in slant depth is very similar to that of
vertical showers in the absence of an electric field. Fig.~\ref{nofield3i} shows the number of electrons as a
function of atmospheric depth for a vertical shower of $10^{16}$ eV and two inclined showers of the same
energy, with zenith angles of 30 and 60 degrees. For inclined showers the slanted atmospheric depth is used, so their evolution should coincide with
a vertical shower but the same depth corresponds to a higher altitude.
The development is the same within error bars. The inclined
showers continue to larger atmospheric depth, as the simulation stops when the particles hit ground level.
As is the case with vertical showers, the shower development is only slightly affected by a field of 100 V/cm. Fig.~\ref{smallfield30} shows
the number of electrons and positrons for negatively and positively aligned fields of 100 V/cm for a shower with a 30 degrees
zenith angle. The variations are of the order of the 1 sigma error bars. Fig.~\ref{smallfield60} contains the same
information for a shower with 60 degrees zenith angle. Again, the electric field effect is small.

\begin{figure}[htp]
\centerline{\epsfig{file=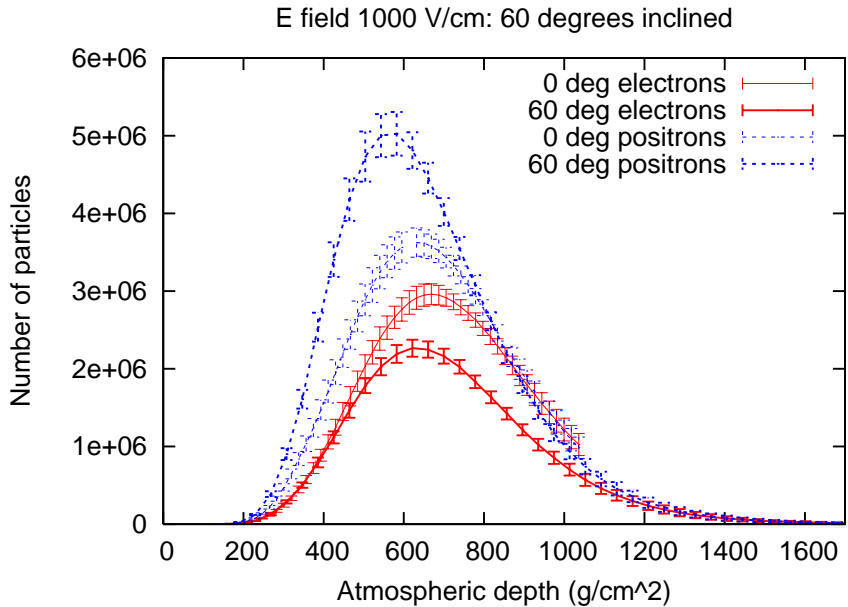 ,width=0.6\textwidth, angle=270}}
\caption{Shower development of a 60 degrees inclined and vertical $10^{16}$~eV shower in a strong positive electric field.} 
\label{incl1000}
\end{figure}

Fig.~\ref{incl1000} shows the shower development for vertical and inclined showers in a strong, positively directed, vertical 
field of 1000 V/cm (accelerating the positrons). The positive charge excess becomes even more pronounced for the 60 degrees
shower. The reason for this change is that for an inclined shower the atmosphere is less dense at the the same slant depth. In a region of low density the energy loss due
to collisions is smaller and acceleration in the electric field more efficient.
 
\begin{figure}[htp]
\centerline{\epsfig{file=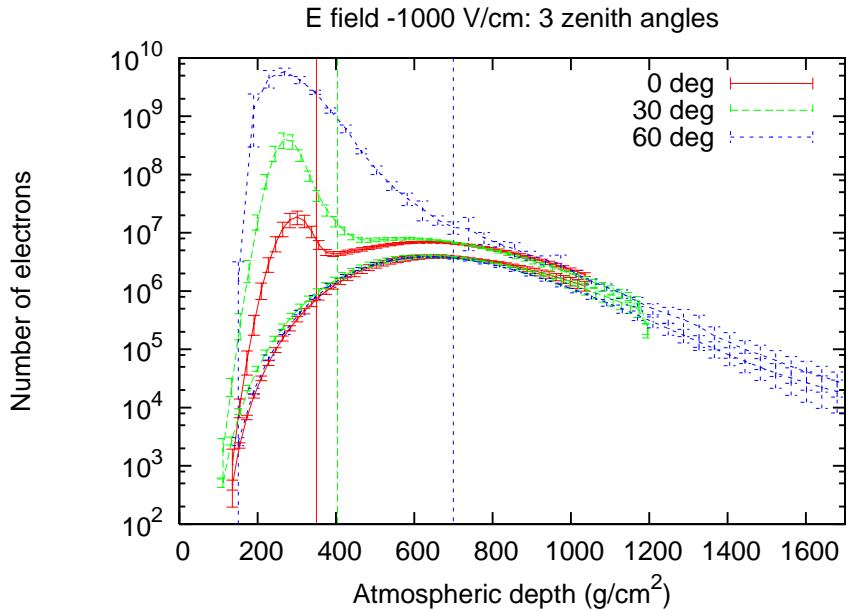 ,width=0.6\textwidth, angle=270}}
\caption{Number of electrons as a function of atmospheric depth for a vertical and 2 inclined showers of $10^{16}$~eV. For each shower
the lower line represents the shower development in the absence of a field. The vertical lines mark the slanted depth at which the electric field
equals the threshold field for the different inclinations. } 
\label{incl-1000}
\end{figure}

The same effect but even stronger can be seen in the development of inclined showers in 
a negatively aligned field (see Fig.~\ref{incl-1000}). The slant depth at which the electric field equals the threshold field is larger for inclined showers and the number
of breakdown electrons is increased by more than three orders of magnitude. Also, the breakdown is sustained longer in terms of slant depth.
The breakdown electrons will be deflected in the electric field and will move away from the shower axis. The area that is affected by the
breakdown effect can have a large horizontal extent.

\begin{figure}[htp]
\centerline{\epsfig{file=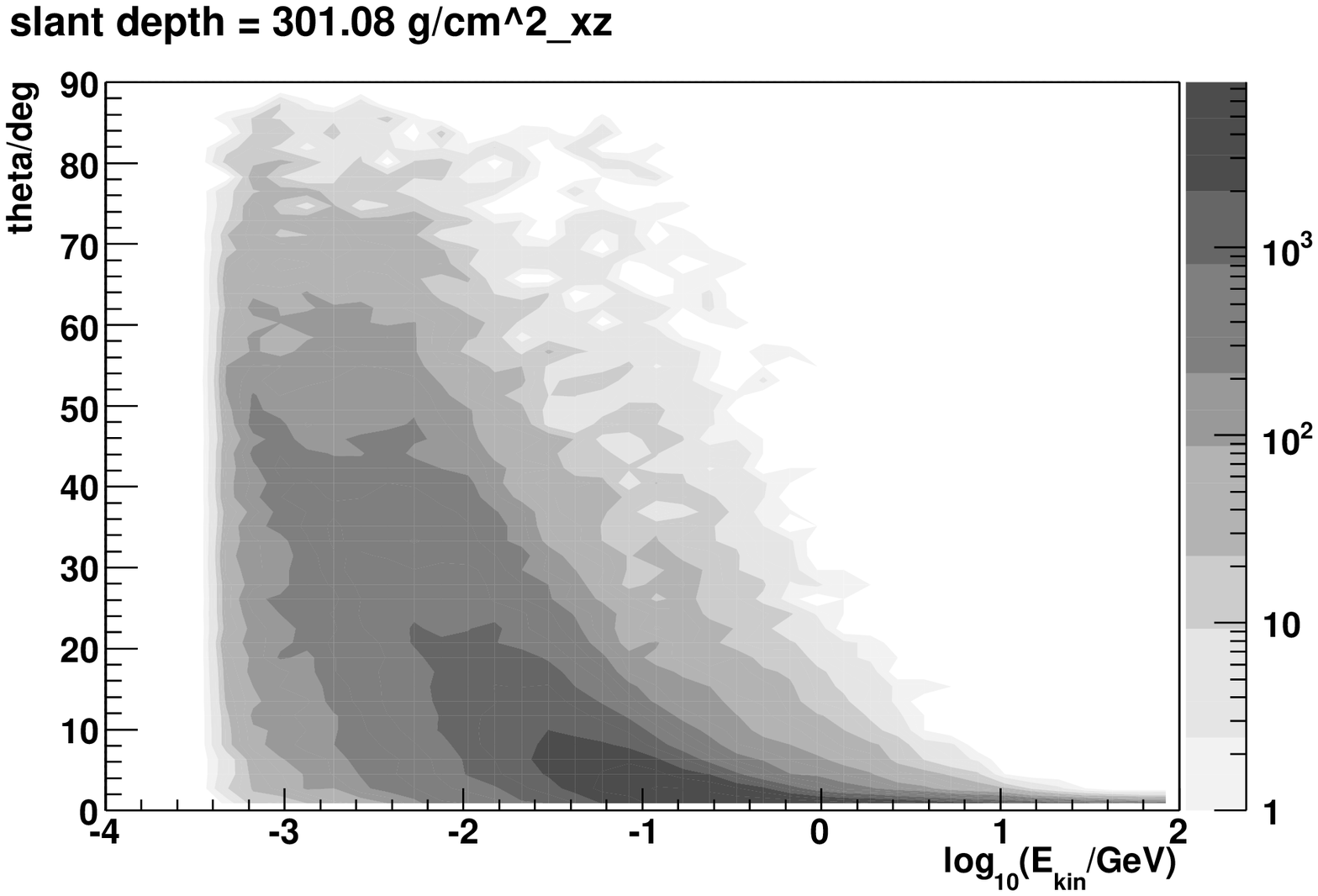 ,width=0.5\textwidth}
            \epsfig{file=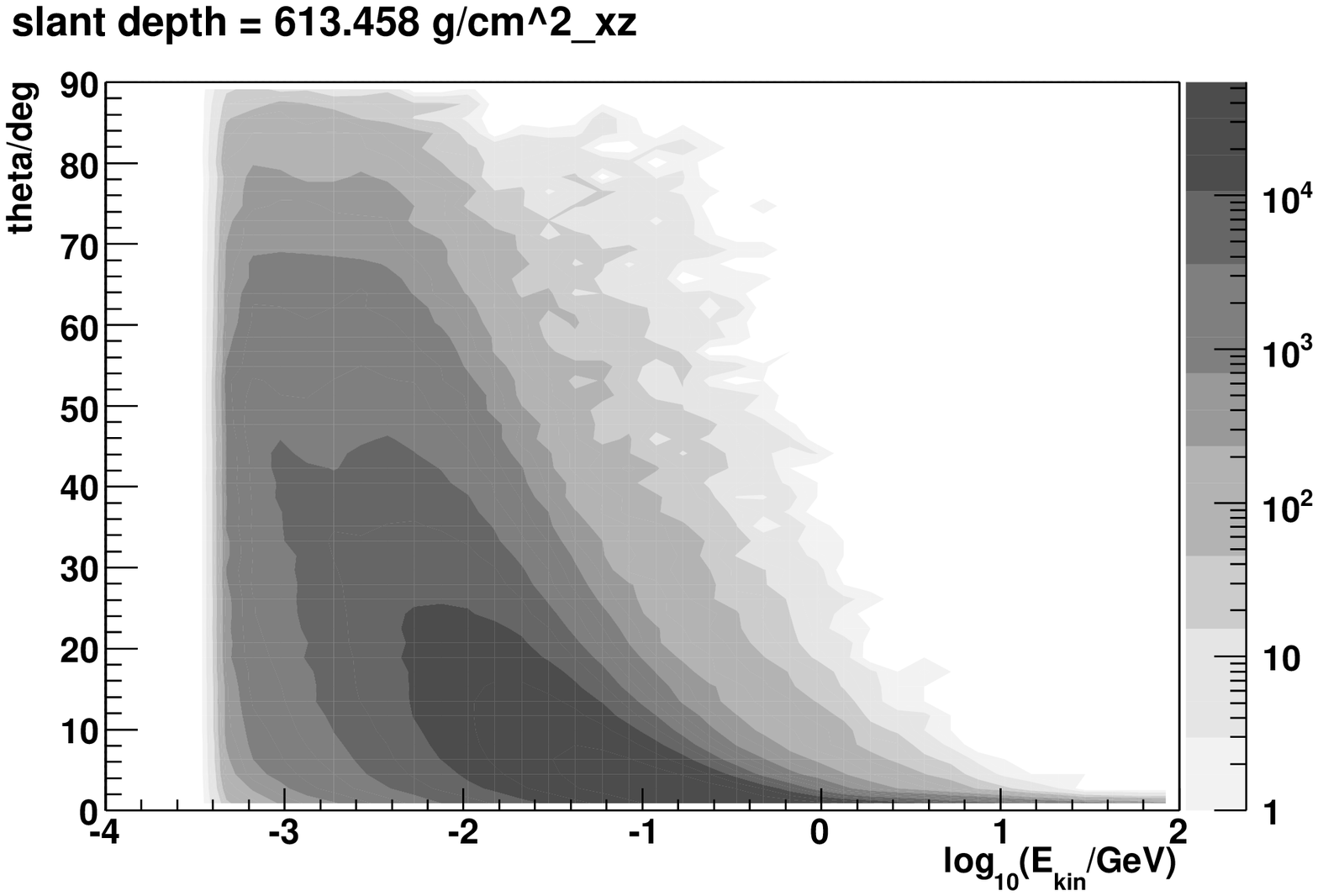 ,width=0.5\textwidth}}
\centerline{\epsfig{file=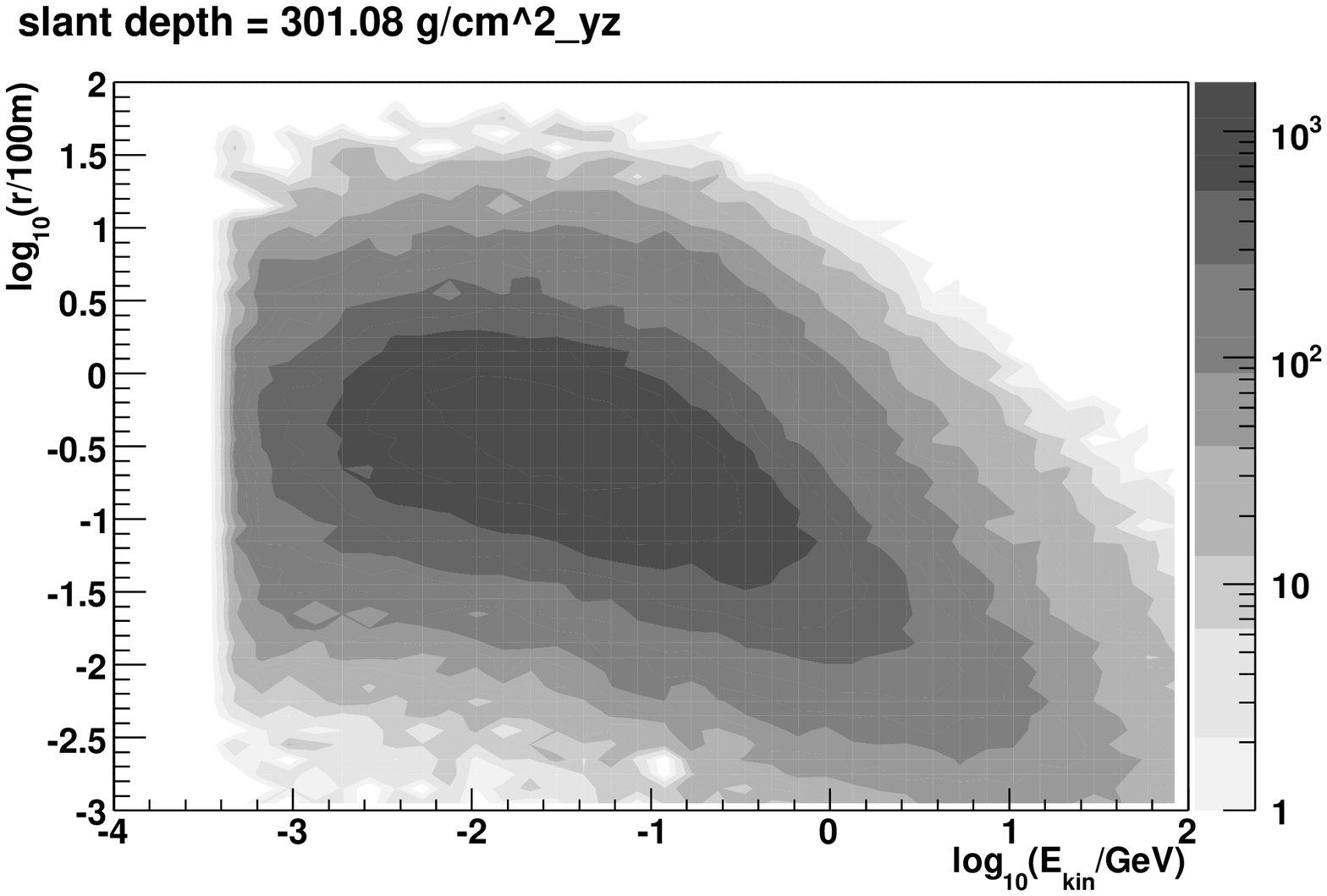 ,width=0.5\textwidth}
            \epsfig{file=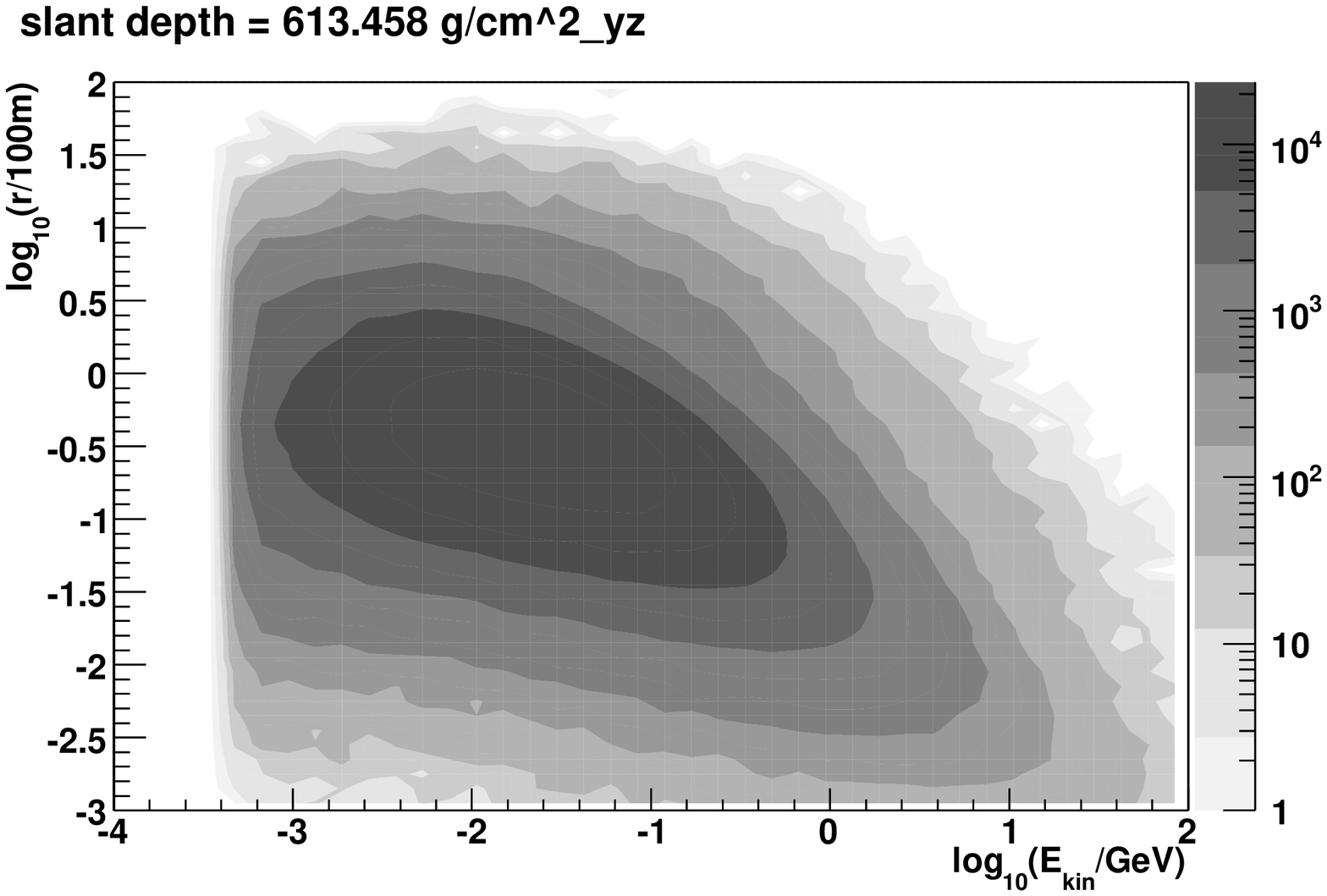 ,width=0.5\textwidth}}
\centerline{\epsfig{file=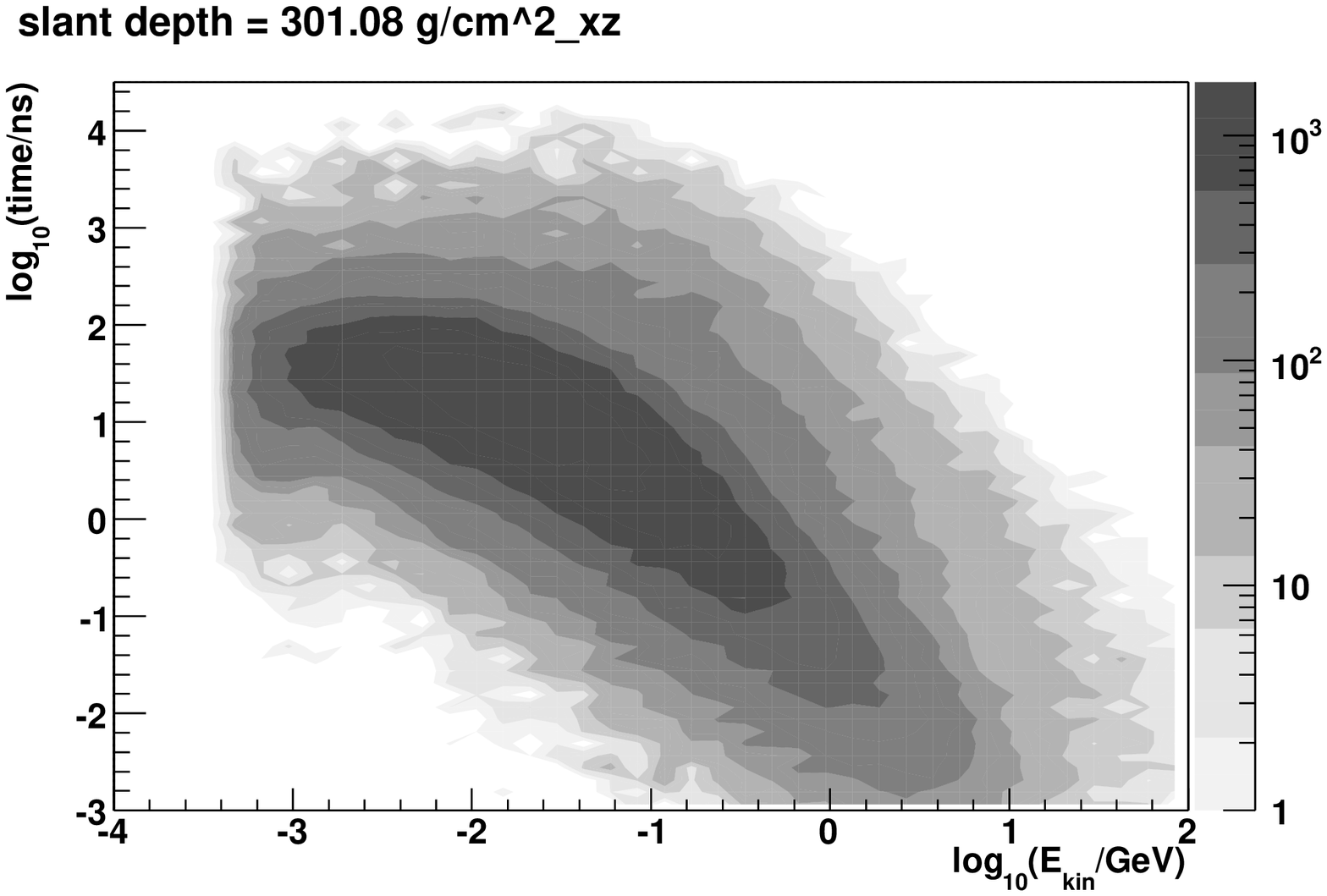 ,width=0.5\textwidth}
            \epsfig{file=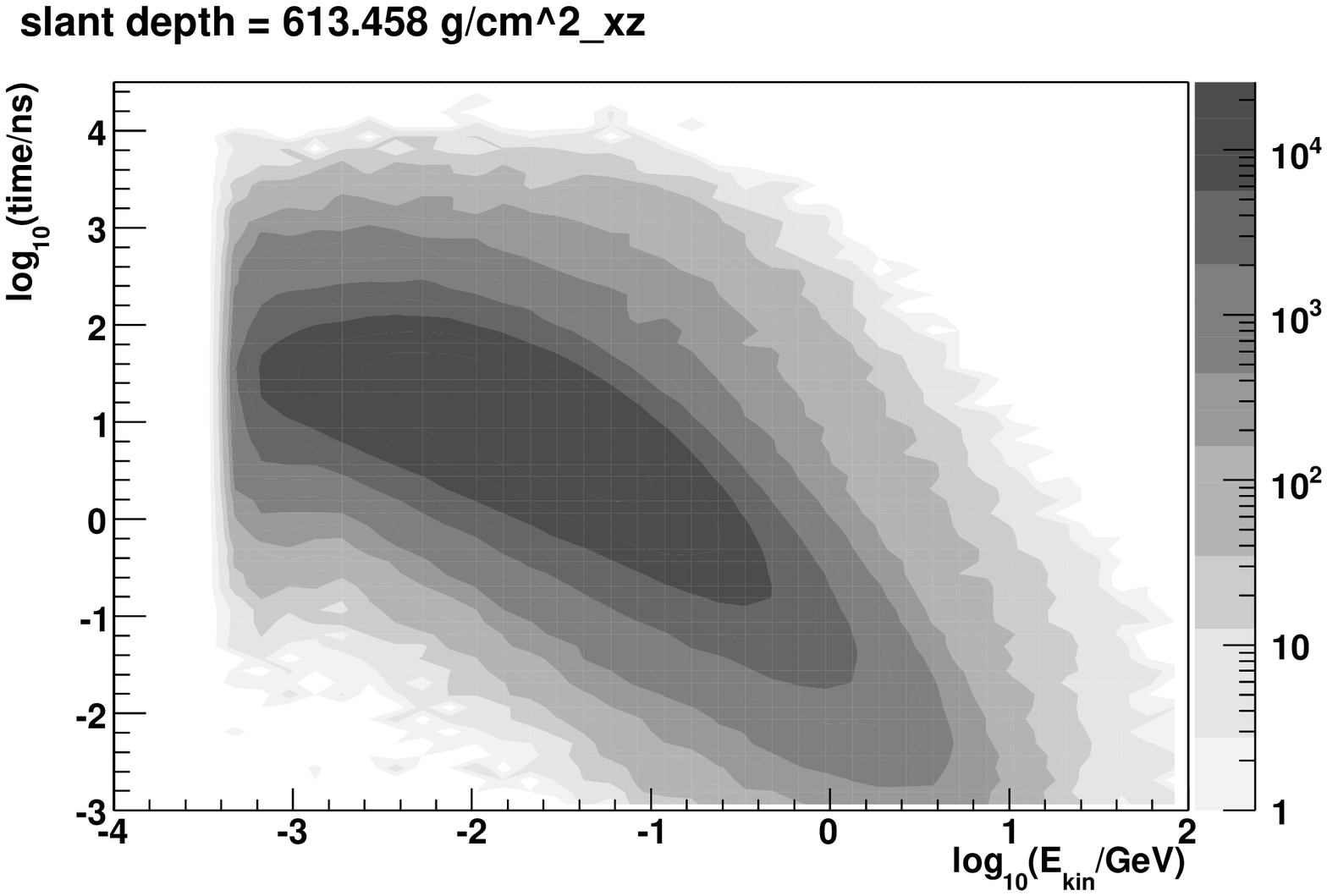 ,width=0.5\textwidth}}
\caption{Electron distributions for a vertical shower with no electric field. Top to bottom: angle w.r.t. shower axis,
distance to shower axis and delay time vs. energy in each case. Left: runaway region, right: shower maximum.} 
\label{s257}
\end{figure}

\begin{figure}[htp]
\centerline{\epsfig{file=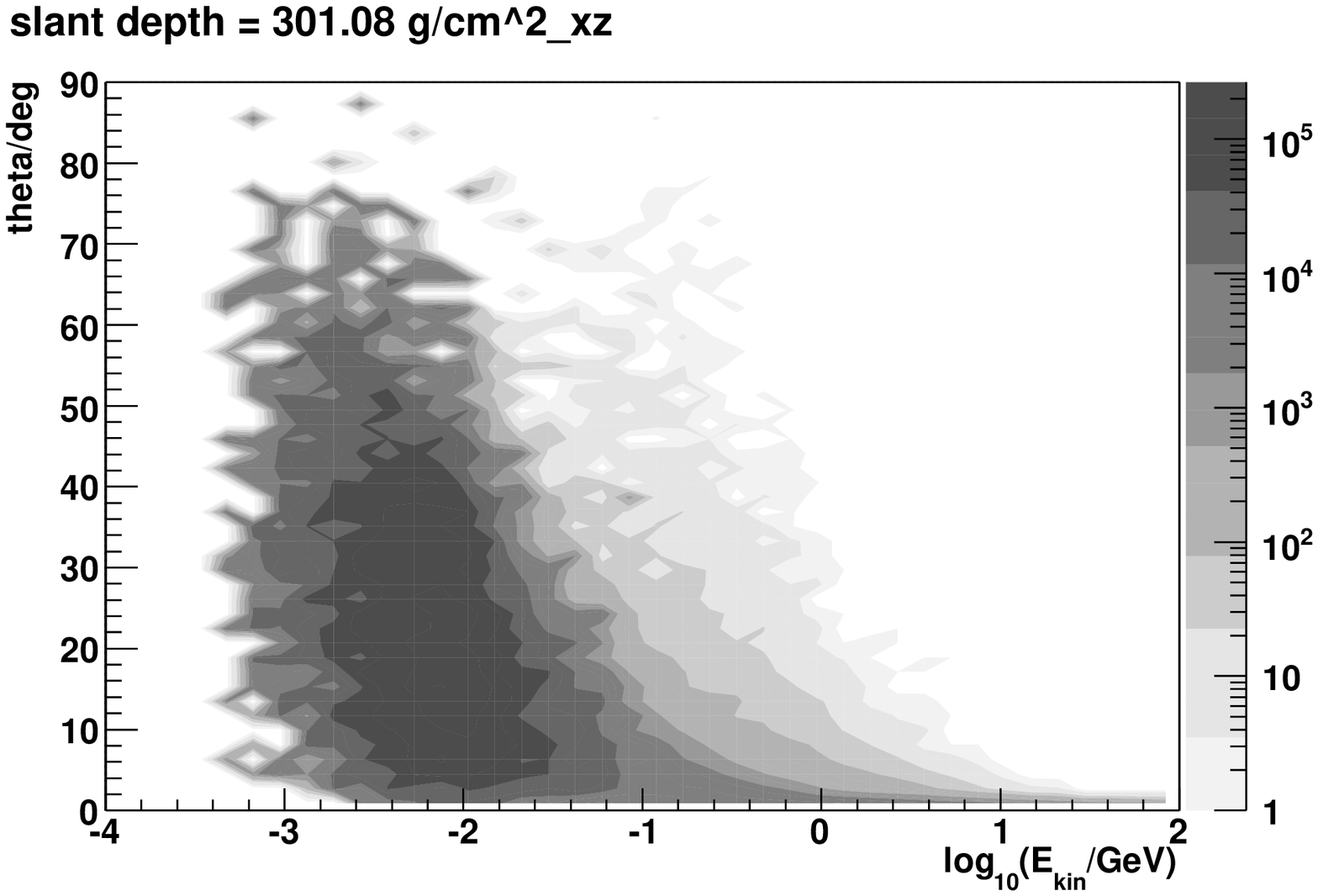 ,width=0.5\textwidth}
            \epsfig{file=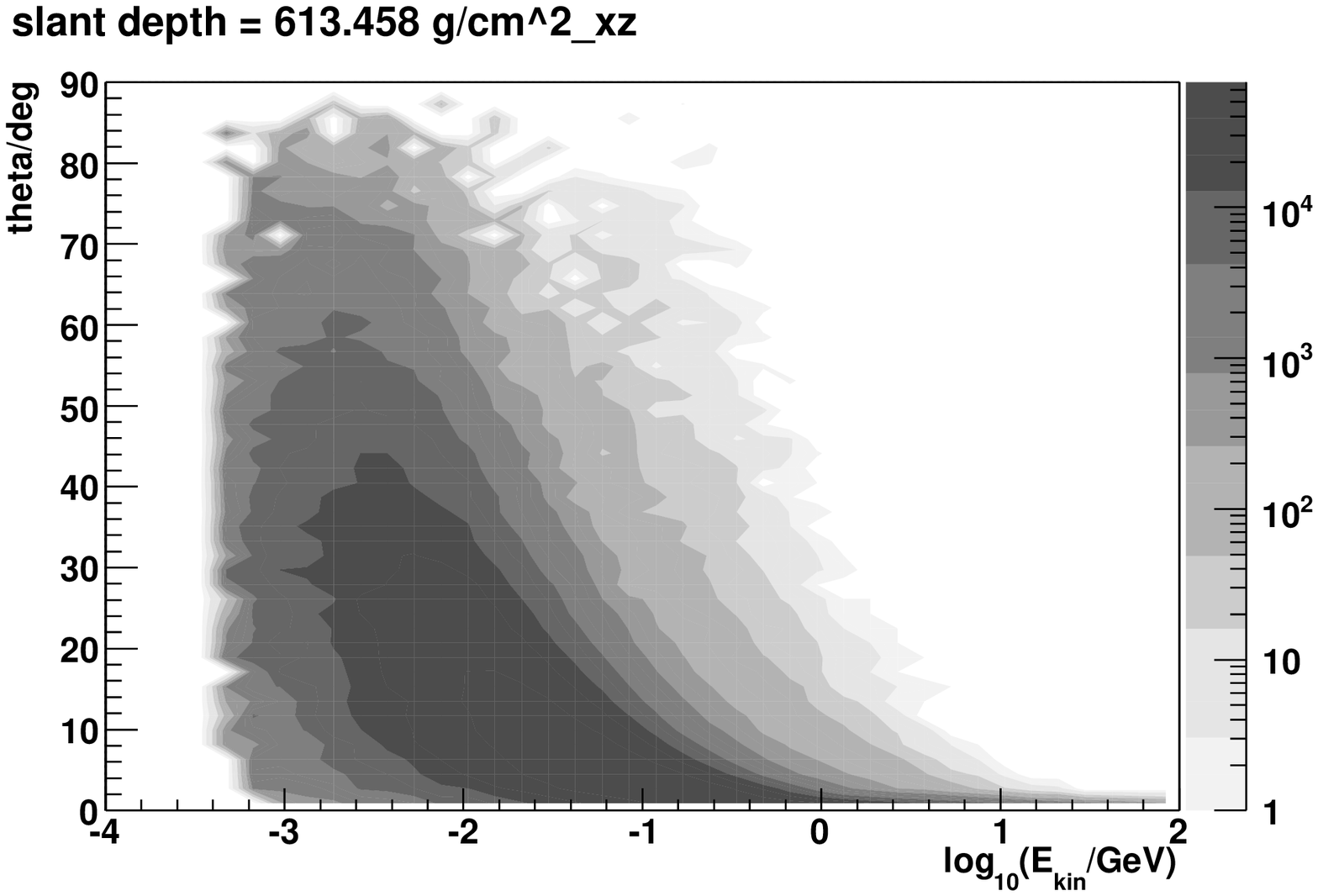 ,width=0.5\textwidth}}
\centerline{\epsfig{file=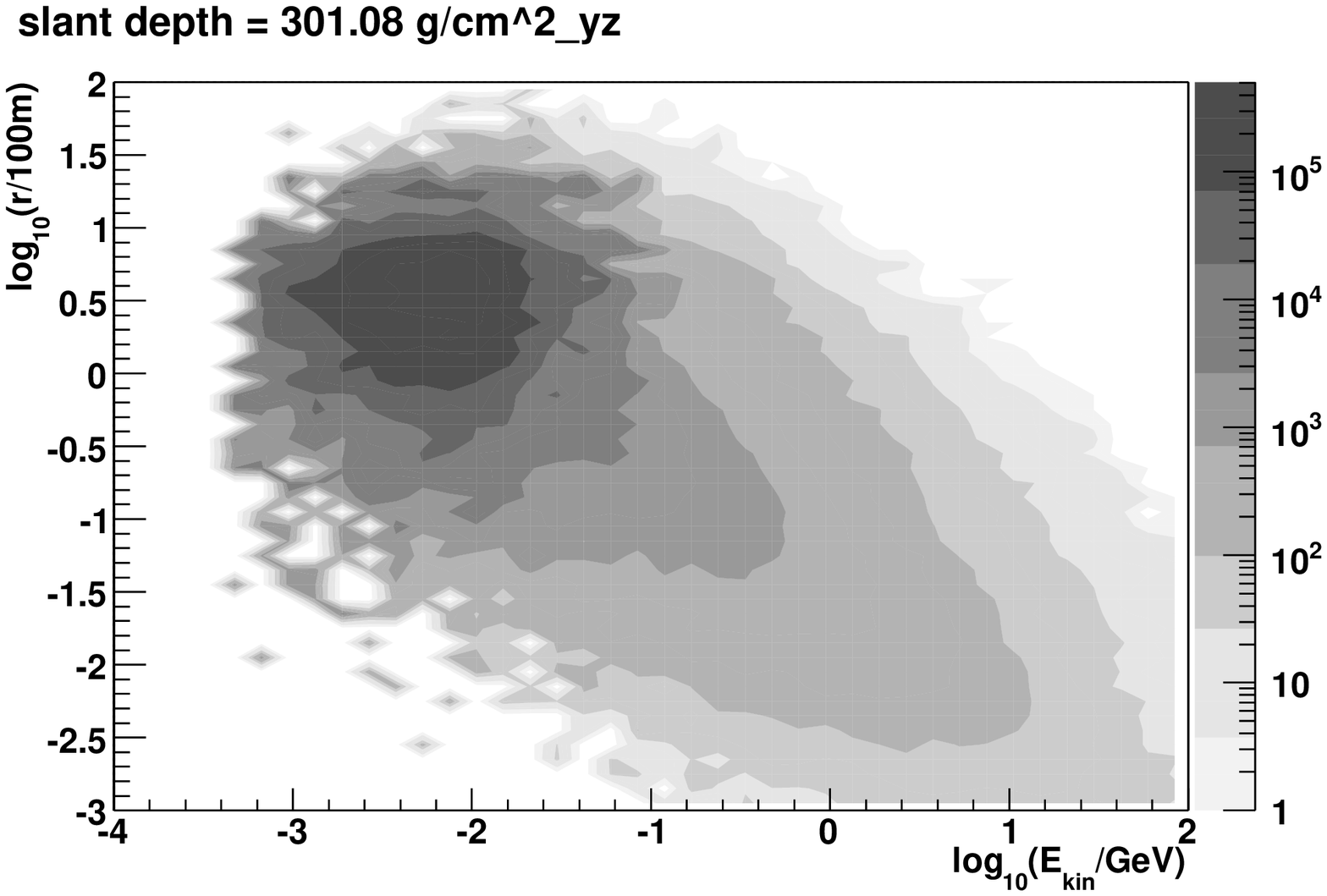 ,width=0.5\textwidth}
            \epsfig{file=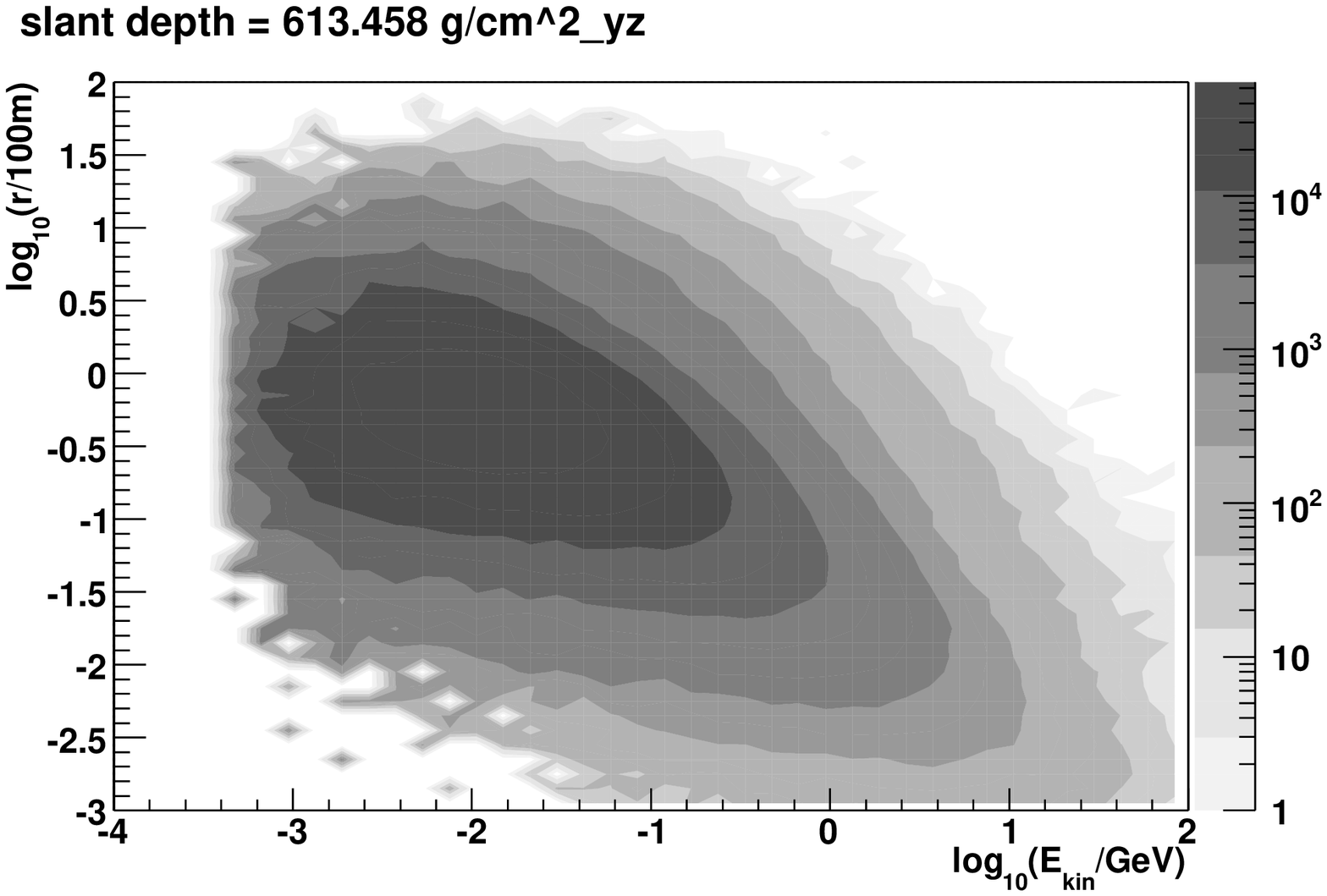 ,width=0.5\textwidth}}
\centerline{\epsfig{file=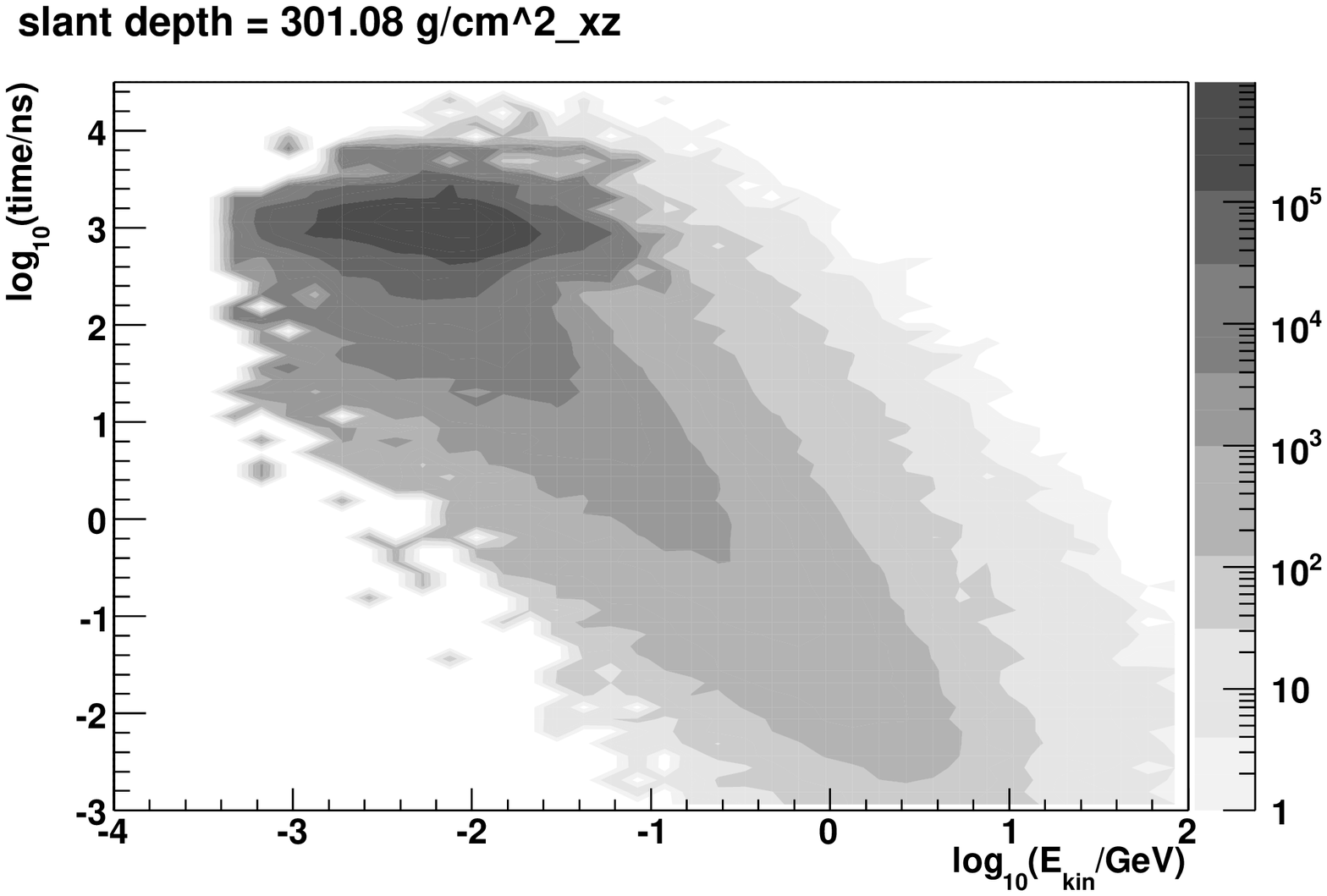 ,width=0.5\textwidth}
            \epsfig{file=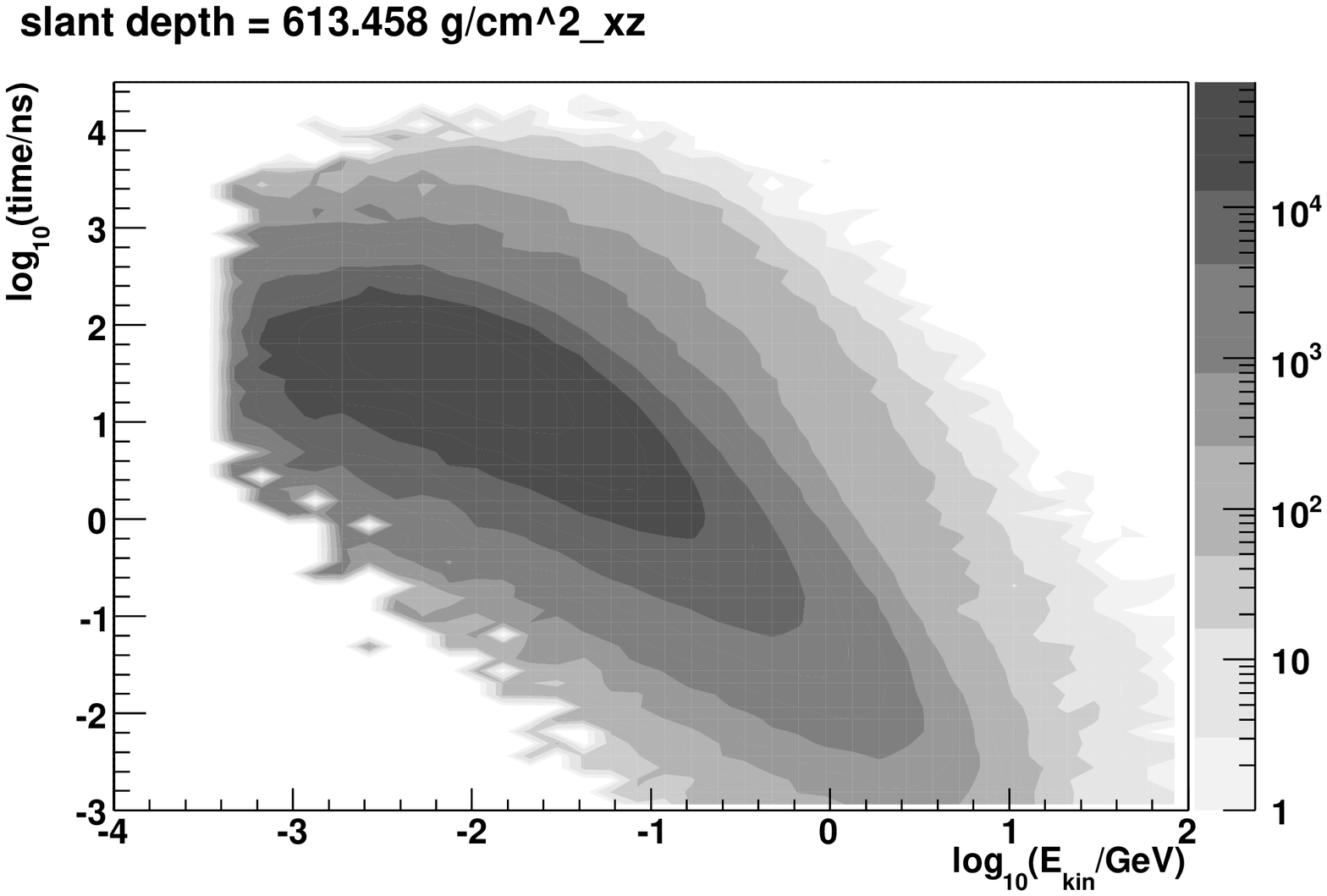 ,width=0.5\textwidth}}
\caption{Electron distributions for a vertical shower, E=-1000 V/cm. Top to bottom: angle w.r.t. shower axis,
distance to shower axis and delay time vs. energy in each case. Left: runaway region, right: shower maximum.} 
\label{s503}
\end{figure}

\begin{figure}[htp]
\centerline{\epsfig{file=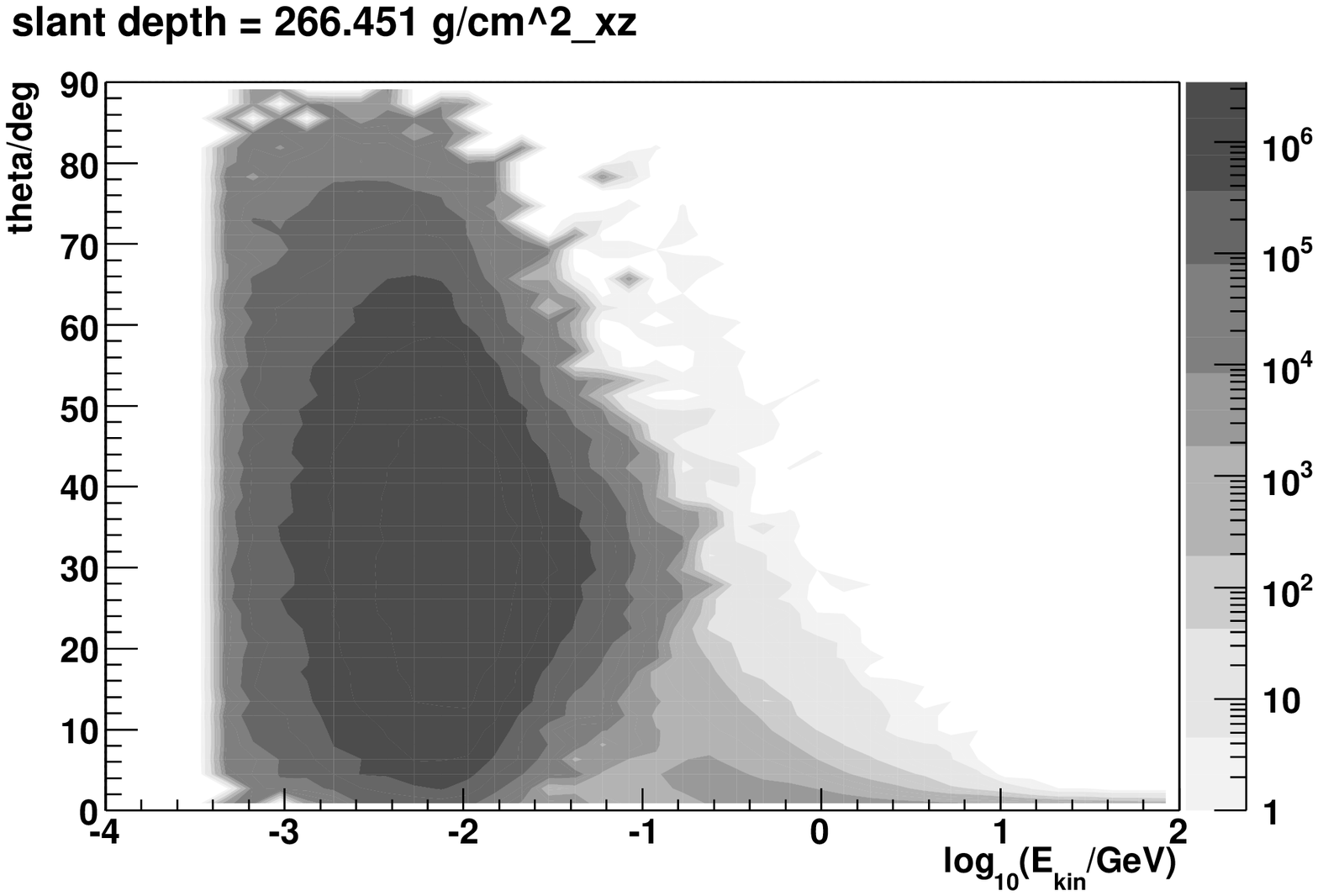 ,width=0.5\textwidth}
            \epsfig{file=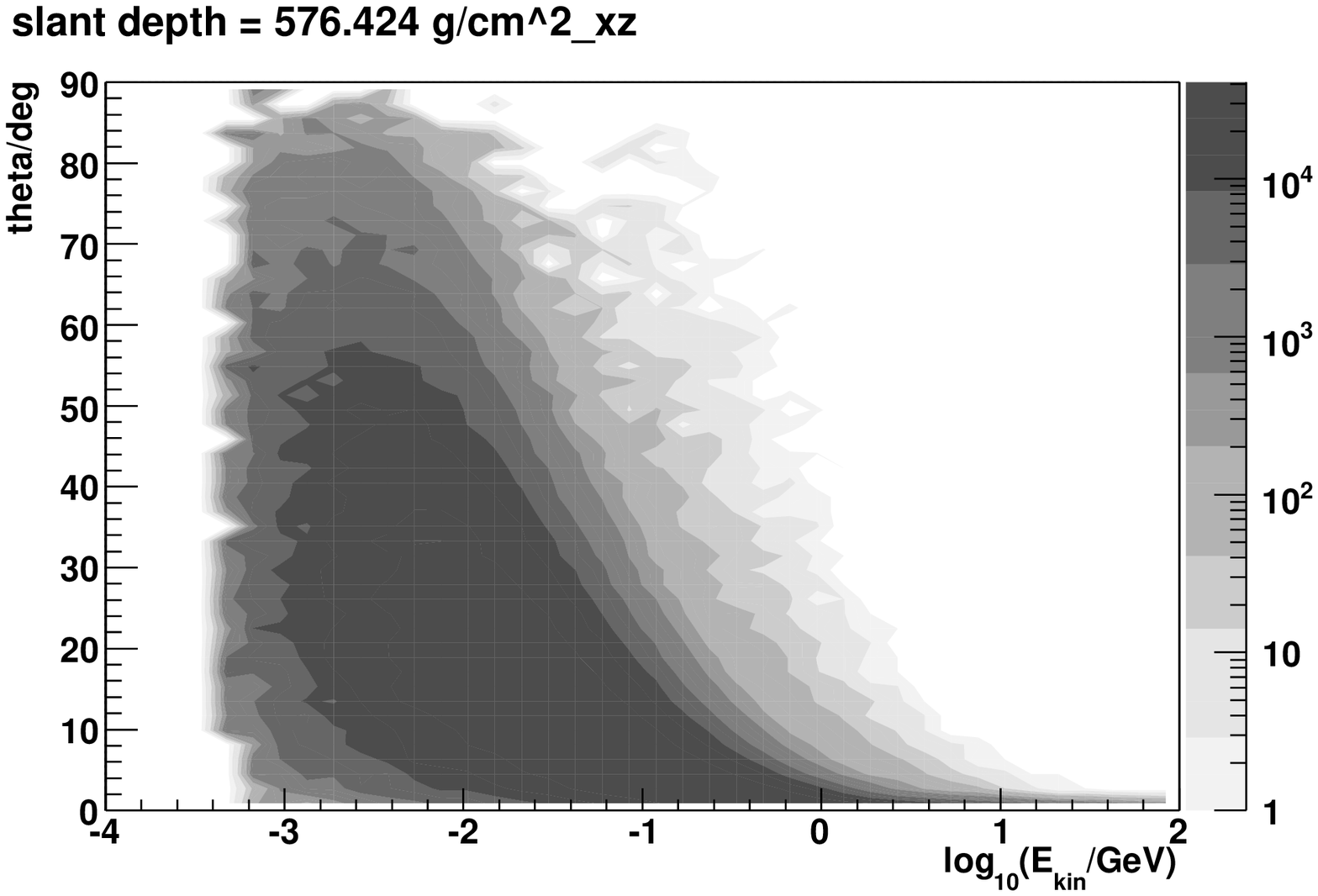 ,width=0.5\textwidth}}
\centerline{\epsfig{file=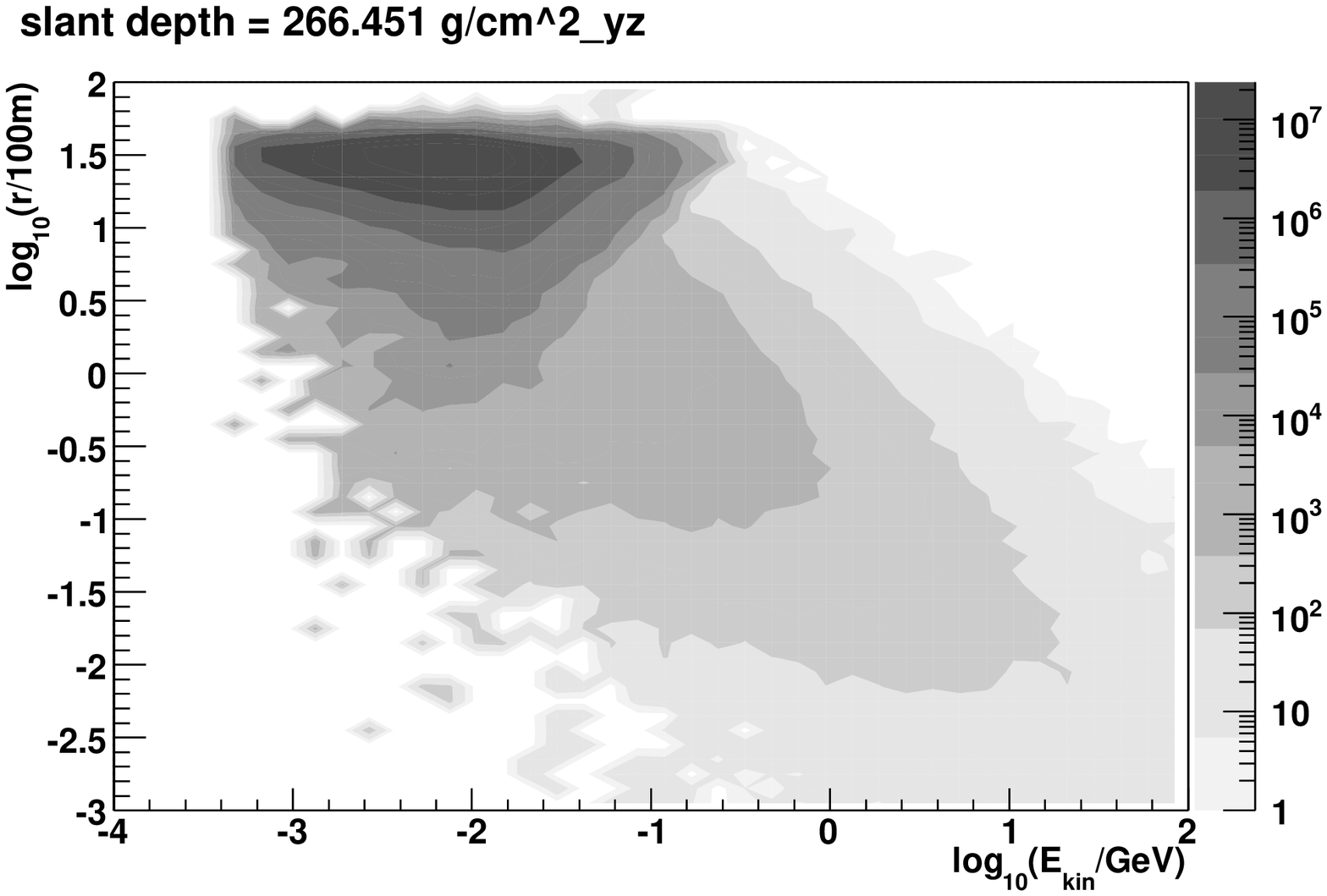 ,width=0.5\textwidth}
            \epsfig{file=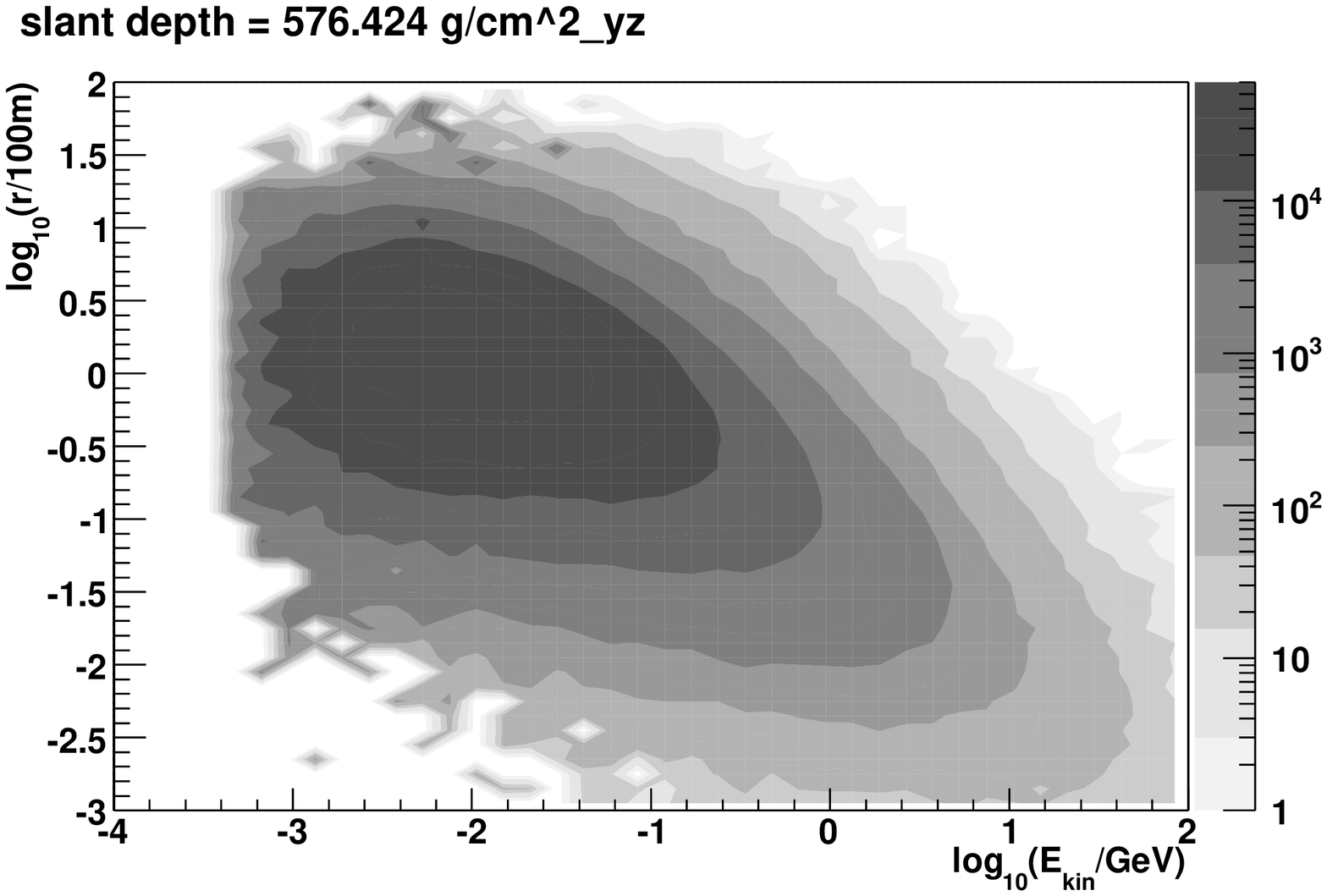 ,width=0.5\textwidth}}
\centerline{\epsfig{file=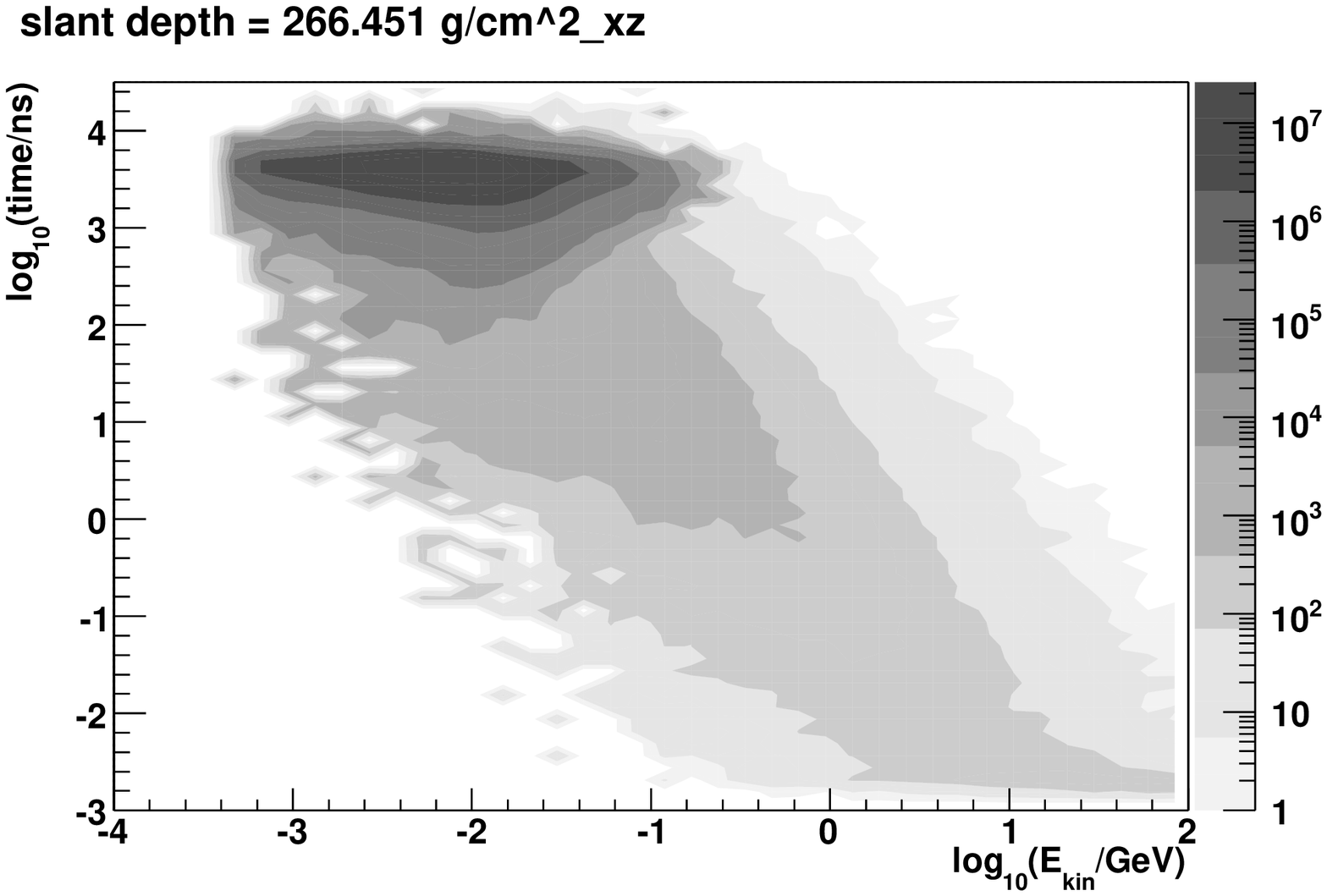 ,width=0.5\textwidth}
            \epsfig{file=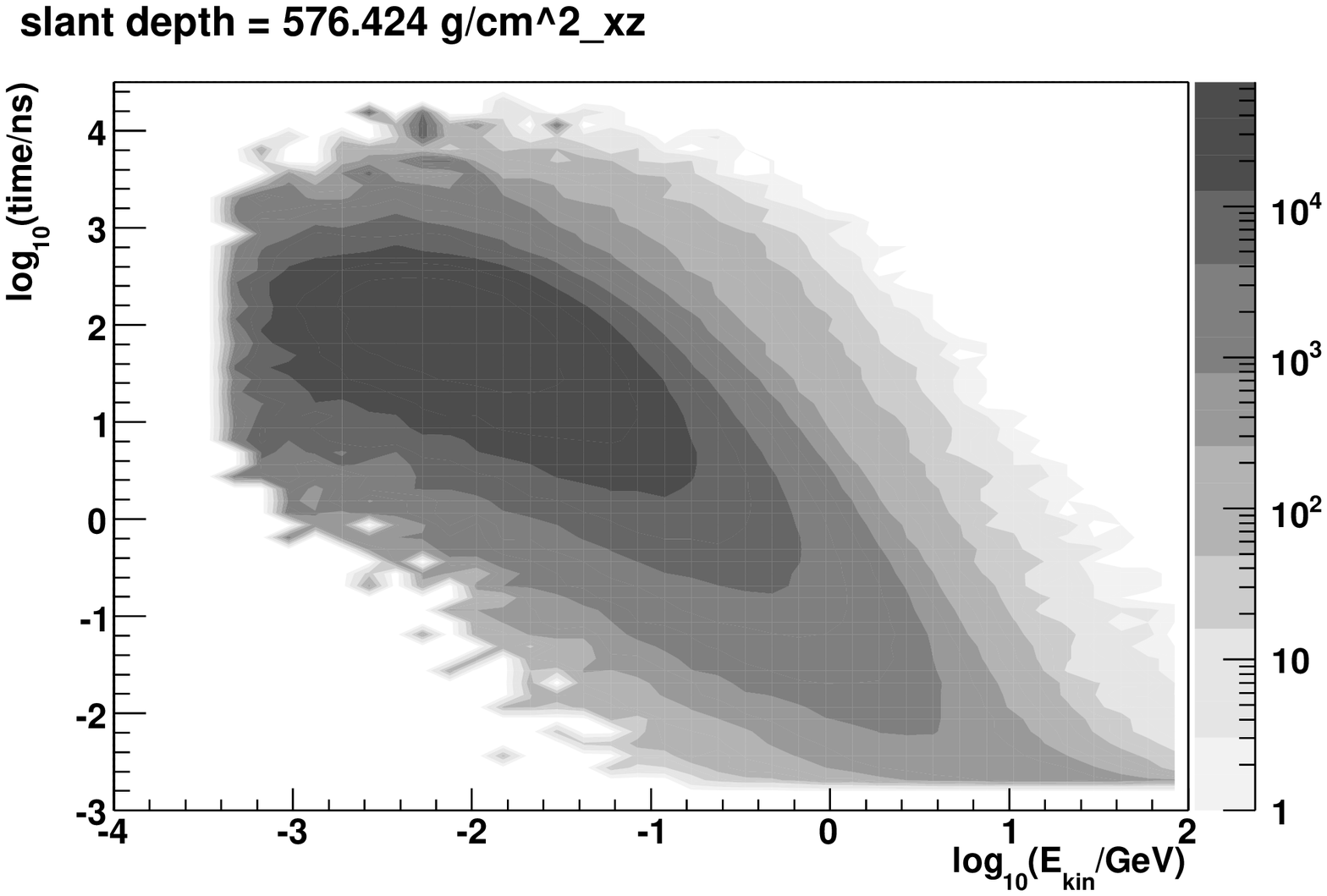 ,width=0.5\textwidth}}
\caption{Electron distributions for z=30 degrees, E=-1000 V/cm. Top to bottom: angle w.r.t. shower axis,
distance to shower axis and delay time vs. energy in each case. Left: runaway region, right: shower maximum.} 
\label{s500500}
\end{figure}

\begin{figure}[htp]
\centerline{\epsfig{file=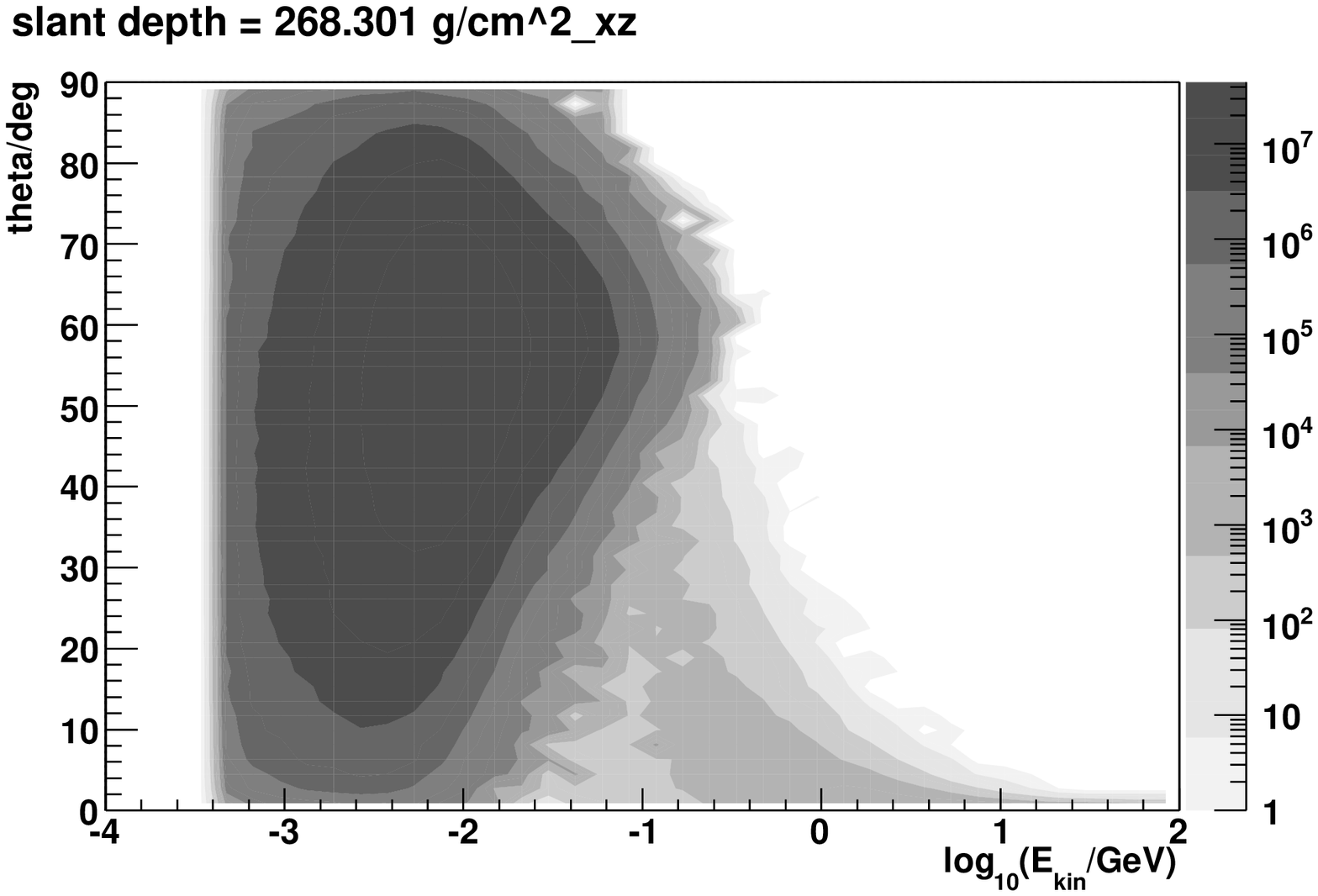 ,width=0.5\textwidth}
            \epsfig{file=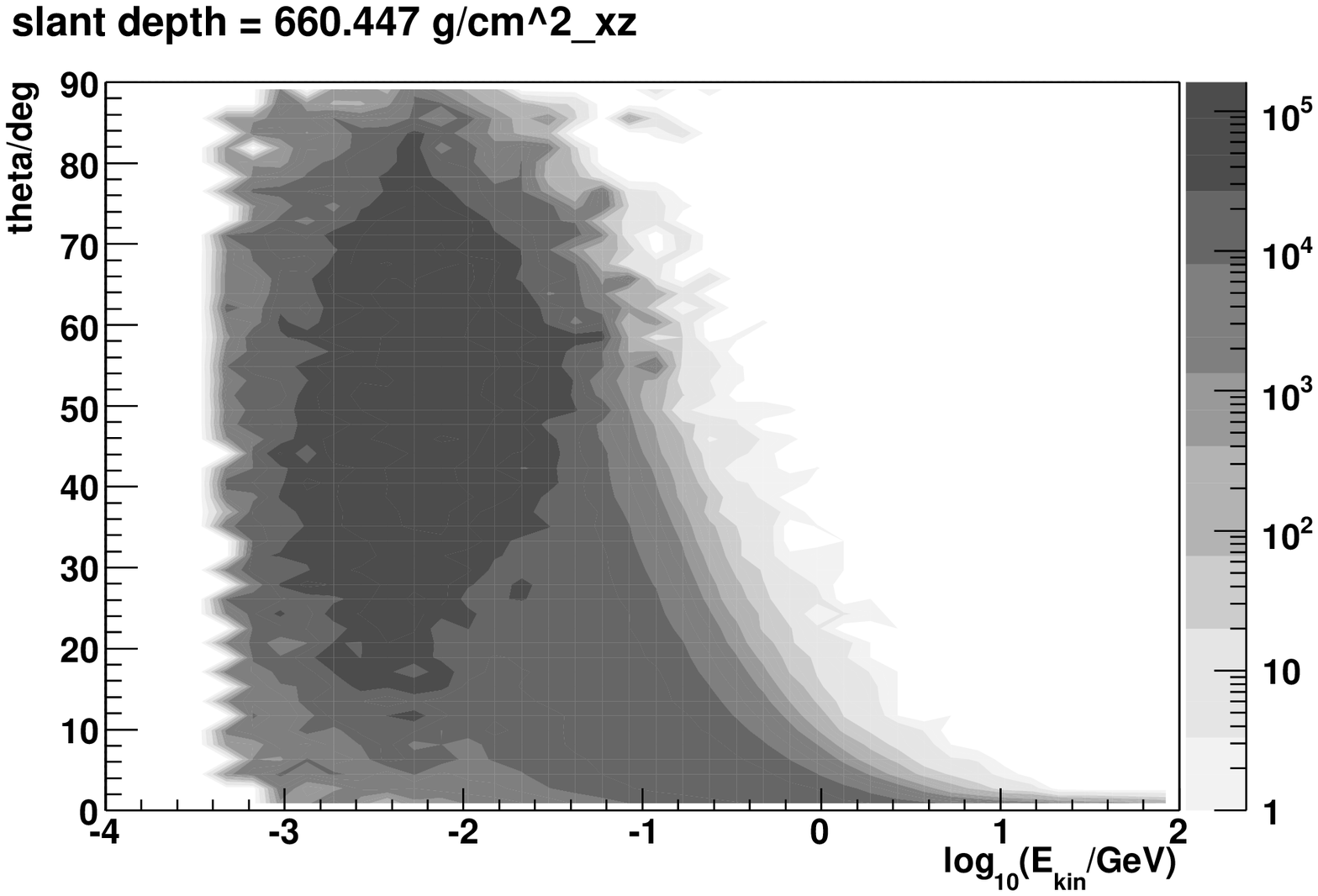 ,width=0.5\textwidth}}
\centerline{\epsfig{file=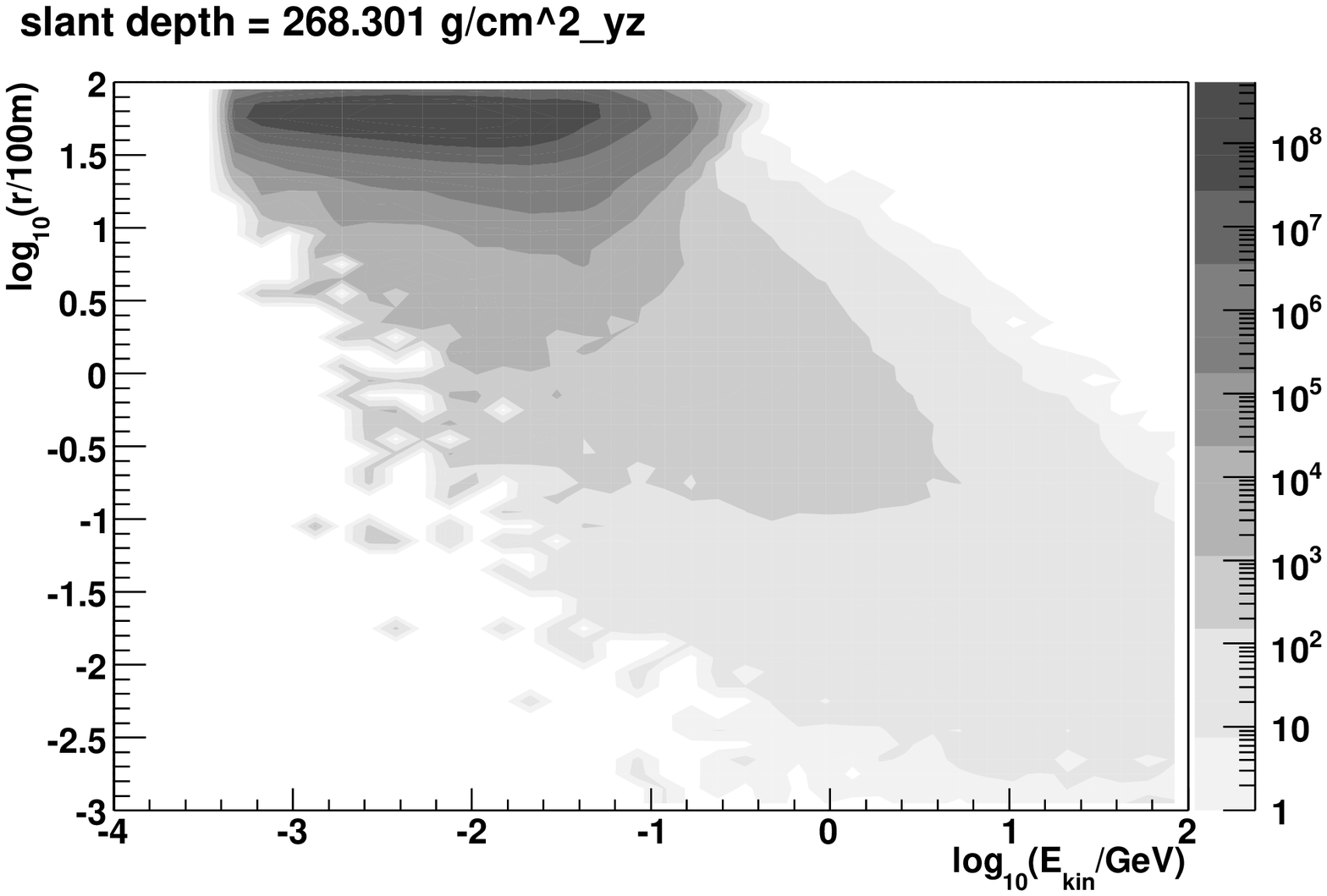 ,width=0.5\textwidth}
            \epsfig{file=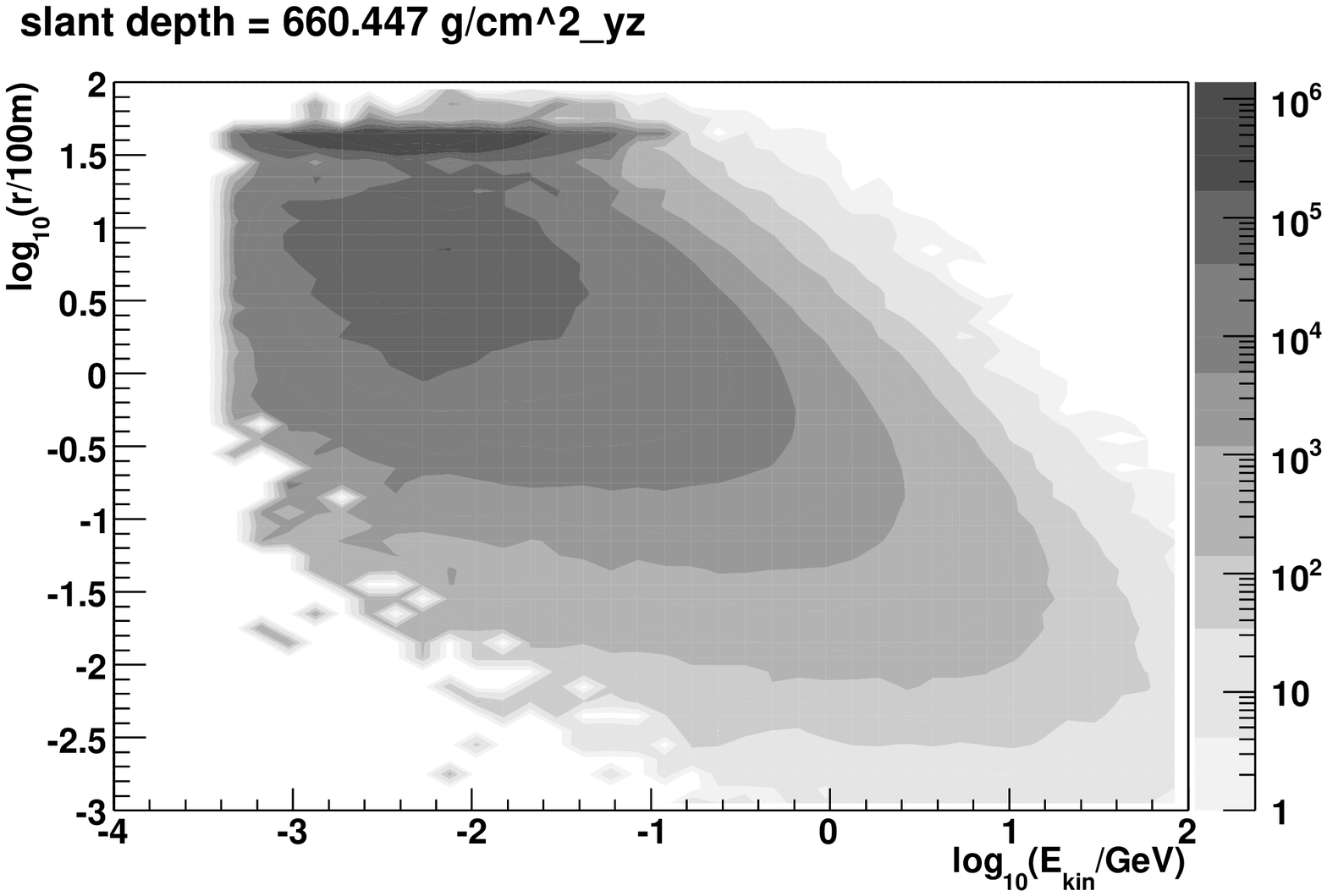 ,width=0.5\textwidth}}
\centerline{\epsfig{file=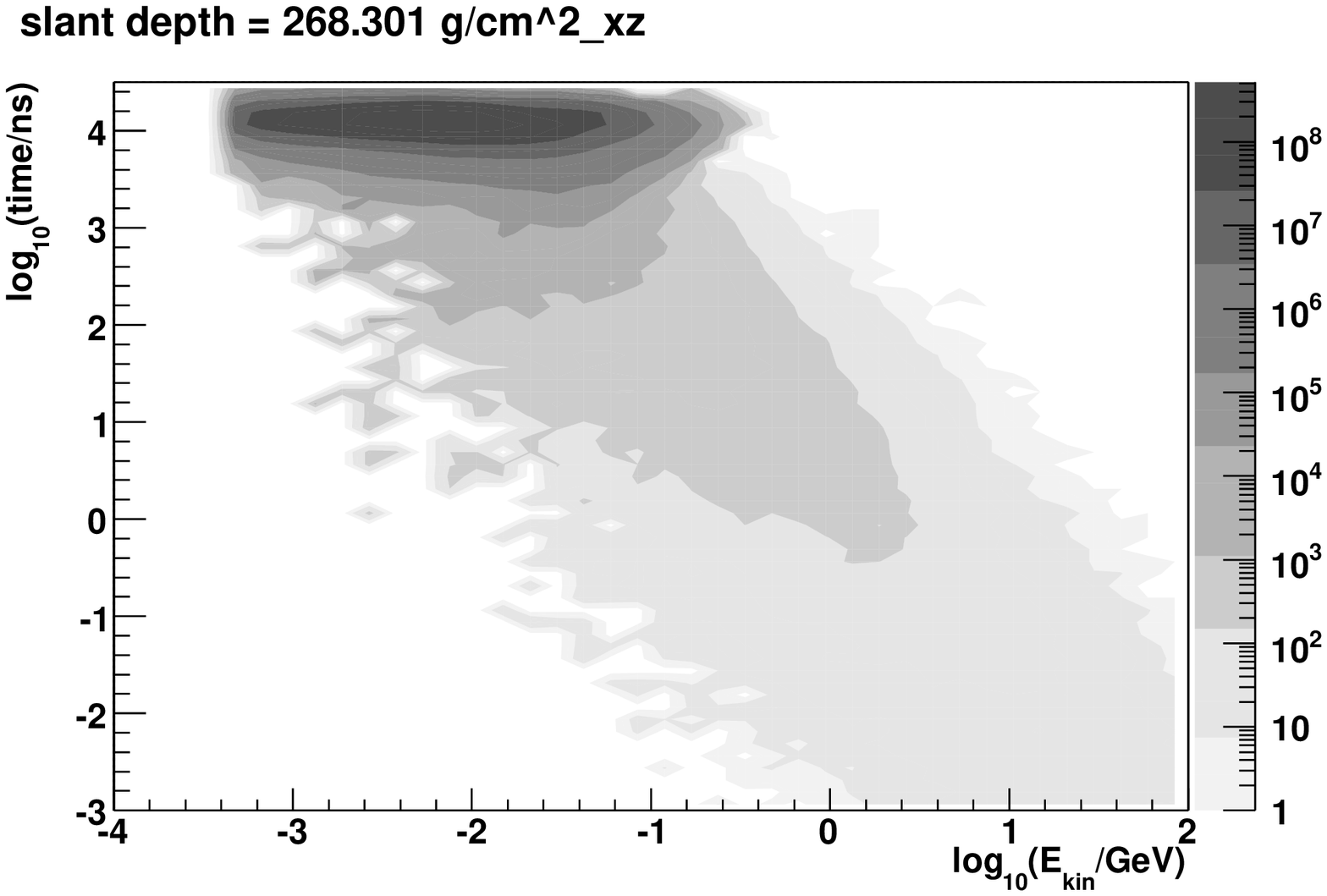 ,width=0.5\textwidth}
            \epsfig{file=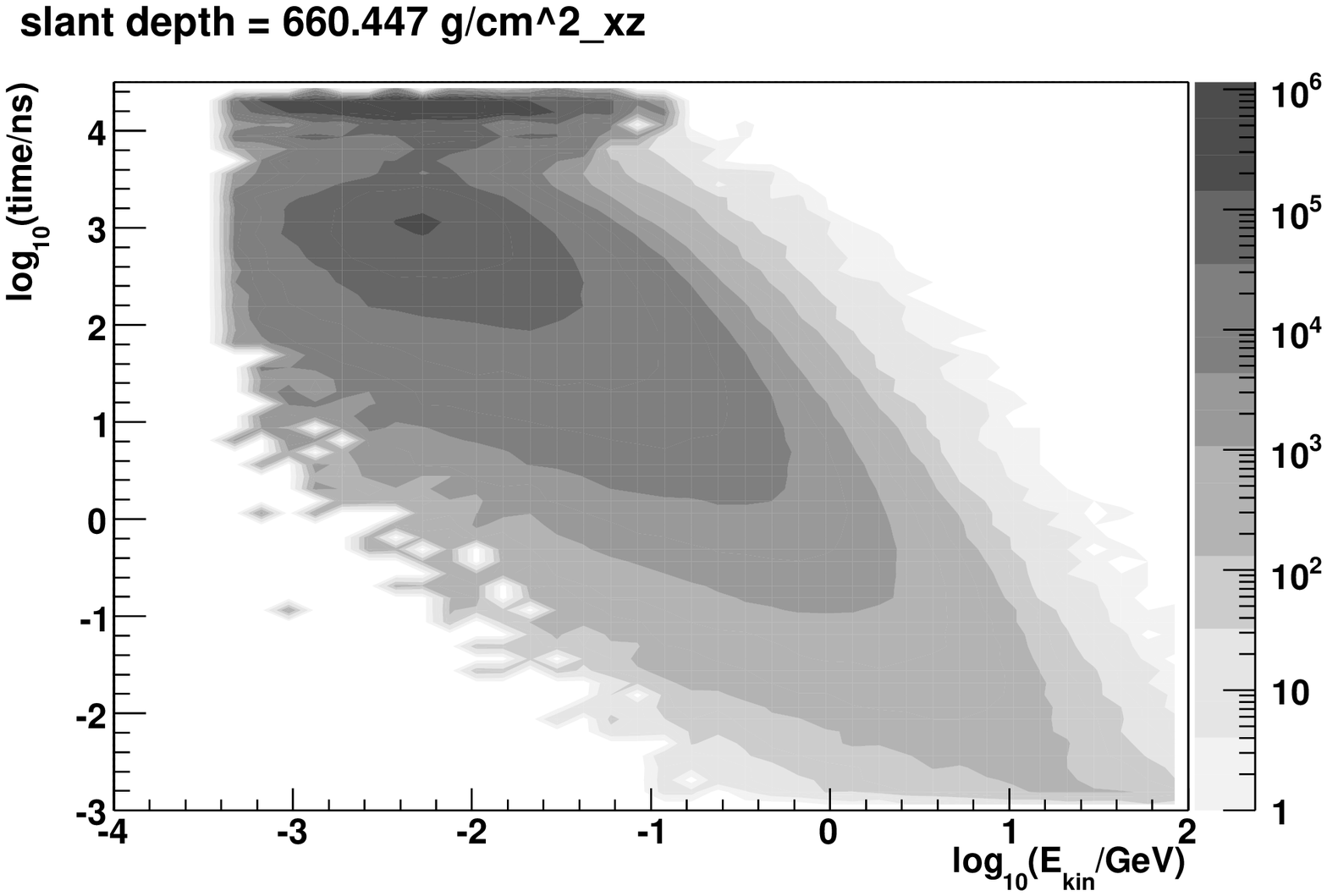 ,width=0.5\textwidth}}
\caption{Electron distributions for z=60 degrees, E=-1000 V/cm. Top to bottom: angle w.r.t. shower axis,
distance to shower axis and delay time vs. energy in each case. Left: runaway region, right: shower maximum.}
\label{s700505}
\end{figure}

Figs.~\ref{s257} through \ref{s700505} show electron distributions for different showers. For each figure the left column corresponds to
the breakdown region and the right column to the shower maximum. From top to bottom, plotted along the vertical axis are respectively, angle of the particle
momentum with respect to the shower axis, distance from the shower axis, and delay time w.r.t.\ a hypothetical shower front
travelling at the speed of light. Each plot has the particle energy on the horizontal axis. Fig.~\ref{s257} shows a vertical shower in
the absence of an electric field. Of course, no breakdown occurs and the shape of the distributions at high altitude is roughly the same 
as the shape of the distributions at shower maximum. Fig.~\ref{s503} shows the distributions for a vertical shower in a negative electric field of 1000 V/cm. 
The characteristics of the 
shower maximum are not
much different from those of the shower maximum in Fig.~\ref{s257} but in the breakdown region clear differences can be seen. In the first place the number of electrons is about a
factor of 100 higher, but the characteristics of the distribution have also changed. The bulk of
the particles has a lower energy, peaking around 10 MeV and the delay time is larger. These are the characteristics that would be
expected from an electron avalanche made up of electrons that start with low energies and are accelerated in the electric field.

Fig.~\ref{s500500} corresponds to a shower in the same electric field with a zenith angle of 30 degrees. It can now be seen that
in the breakdown region the bulk of the electrons has an angle between 20 and 40 degrees with the shower axis. The reason for this
is that the low energy electrons are deflected into the direction of the electric field (i.e.\ vertical in the case studied here). For the same reason the distance to the
shower axis and the delay time increases. At the shower maximum the distributions return roughly to the shapes of the vertical shower distributions.

For the shower with 60 degrees zenith angle the distributions are plotted in Fig.~\ref{s700505}. Now, the shower 
maximum is located at such an altitude that the breakdown has not yet ended. In the left column an
electron avalanche is seen of which the particles have an even larger angle with respect to the shower axis and also larger distances and
delay times. In the right column, two contributions can be distinguished: breakdown electrons having the same characteristics as in
the left column and shower electrons that are produced via pion decay and feature a more traditional distribution.


\section{Discussion}
\label{sec:dis}
The electric fields we have used in our simulation are completely homogeneous and not realistic. We used these simple fields to clarify
the effects the electric field can have on the shower development and make it easier to compare the results to rough analytic
limits. Inside a thunderstorm, the electric field will strongly change in strength and direction with altitude. The effects
described in this paper will only occur locally at the region where the field is strongest. 

Electric field measurements inside thunderstorms are difficult to perform because of their violent nature. Marshall et al. \cite{MMR95} present soundings of the vertical component of the electric field measured with balloon-borne field meters. The strength of the field generally fluctuates, but rarely exceeds the breakeven field strength. the polarity can change multiple times. Experiments typically measure fields integrated over a few tens of meters and a few seconds. It is uncertain if stronger field fluctuation exist on smaller spatial or temporal scales. 

For air shower experiments this means that the particle count at detector level does not change strongly, except when the experiment is very close to
or inside the thundercloud. The observed $\sim 10-15\%$ increase in electron count at the high 
altitude experiment EAS-TOP \cite{eastop}
during thunderstorms can be
explained by a nearby electric field with an order of magnitude of 1000 V/cm. The additional particles are either electrons
produced in a breakdown process, or low energy shower electrons and positrons that have gained sufficient energy in the
background electric field to be above the detection threshold.  
The same study also reports an increase in air shower detection rate. This can be the result of an increase in the number of
electrons and positrons in the front of showers which are, in fair weather conditions, below the detection threshold 
Alternatively, it can be due to the deflection of breakdown electrons in the direction of the electric field coming from showers that would have otherwise missed
the detector. Indeed, during the period of increased detection rate, the air showers cluster around a specific (near-vertical) arrival direction. 

The radio pulse of an air shower can be severely affected by strong electric fields. All shower electrons and positrons contribute to the emission
and the largest contribution to the pulse comes from the particles at and just before the shower maximum \cite{REAS2}. When the shower maximum is inside a strong
field region the radio pulse will not only be altered because the energy distribution of the electrons and positrons has changed, but also because
the acceleration itself produces additional radiation. We will study the combination of these effects in a future paper using the radio
simulation code REAS2 \cite{REAS2}. 
The growth and decay of the current that is produced by runaway electrons, results in strong radio emission (see e.g.~\cite{G02,T05,Dw09}).

The intensity of emitted fluorescence light is dependent
on the energy of the electromagnetic shower and can in principle be influenced by atmospheric electric fields. It is,
however, unlikely that strong fields exist outside clouds, where such fluorescence effects can be observed.  

We used CORSIKA without the UPWARD option, so the upward going
particles are not tracked, which have a zenith angle of more than
90 degrees.  The electron
avalanches that are observed in negative field simulations may also occur in the upward direction for positive fields, but are
discarded by CORSIKA. Upward going particles may produce new pairs via bremsstrahlung and pair production, of which one particle
is accelerated downwards again. It is shown by Dwyer \cite{Dw03} that such feedback mechanisms are important in breakdown processes. 
These particles, however, are missed by our simulation. 

The lower energy cutoff in our CORSIKA simulations is set to 0.5 MeV for the electromagnetic shower. This means that electrons below this energy are
discarded before they may be accelerated to higher energies by the electric field. To compensate for collisional energy loss a field of a few times
the breakeven field is needed to accelerate these particles. For our simulation this means the number of electrons above the altitude at which
the field equals the breakeven field may be underestimated.

By discarding particles that have low energies or propagate in the upward direction, the simulations cannot be expected to give a complete description of the runaway breakdown process. The inclusion of lower energy particles will lead to a much larger total number of runaway electrons because of exponential growth. These low energy electrons will be accelerated to relativistic energies by the electric field, but due to elastic scattering the average position of the runaway electrons moves at 0.89c \cite{C06}, considerably slower than the shower particles. For new generations of electrons  produced by upward traveling particles, the delay will be even larger. Moreover, the particles that start out with low energies will not in general be accelerated into the direction of shower propagation. 

This means our simulations are reliable for all particles that are considered to be part of the shower (i.e. located within the shower front, which has a typical thickness of $\sim 1$~m near the shower axis). Furthermore, the simulations show that a shower induces a runaway breakdown process if the background electric field is strong enough. The size of the breakdown is, however, not reliably simulated and probably underestimated. 

Because of its structure, CORSIKA cannot be easily used for simulation of the breakdown itself. 
Inside an area where electron breakdown occurs the electric field will be shielded by the electrons and ions. In effect, the field may be
amplified outside this region. A simulation of the breakdown should have an electric field that varies over time as a function of the charge distribution (see e.g. \cite{Dw05}). CORSIKA traces one particle at a time, making it impossible to include this without major alterations to the code. 
Electrical processes, such as the formation of streamers and lightning initiation, 
are outside the reach of our simulations, but our results provide a strong hint that air showers can trigger such events when they pass through a strong field region 
of a thunderstorm. 

\section{Conclusion}
We have conducted CORSIKA simulations for vertical and inclined showers of several energies inside electric fields with varying
strengths. The evolution of the electromagnetic part of the shower does not change significantly below field strengths of 100
V/cm. For fields of the order of a 1000 V/cm the effect becomes important. Such fields only occur naturally inside thunderclouds. 
The energy distribution can be altered up to energies
of $\sim 1$~GeV. For positive fields this means positrons can outnumber the electrons over a large energy range, resulting in a
positive charge excess. For negative fields electron breakdown is observed for altitudes at which the field is larger than the
breakeven field and is most efficient when the field is larger than the threshold field. This electron avalanche is directed antiparallel to the electric field and can geometrically be detached from the shower.
The electric field effect on the shower evolution is local in the sense that in a low field region a shower that has traversed a
high field region is not much different from a shower that has not. For air showers that traverse thunderstorms this means they
are generally affected only in the strong field regions of the cloud. Ground based particle detectors will not be very sensitive
to these effects.

The radio signal from an air shower that travels through a strong field region can significantly change. It has been established
experimentally that under thunderstorm conditions the power of the radio pulse may be much larger than anticipated. The order of
magnitude of maximum electric fields that occur in thunderstorms coincide with the field strengths at which our simulations show a
considerable change in particle distribution. It should be noted, however, that for the strength of the radio pulse to increase, the shower
evolution does not necessarily have to change. The direct acceleration of the particles by the electric field produces radiation just like the
magnetic deflection does. The effect of electric fields on the radio emission of air showers will be explored in more detail in a
forthcoming paper.

Air showers can trigger electron avalanches in regions where the electric field exceeds the threshold field. These avalanches may
play a role in lightning initiation. Although electron avalanches can be initiated by any seed electron of sufficient energy in a field exceeding the threshold field, a passing
air shower offers the unique scenario in which a huge number of high energy electrons is injected in a very small time. The possible interaction
between air showers and thunderstorms can be investigated with a hybrid array of particle detectors and radio receivers. In
principle radio antennas can pick up the signal of both air showers and electrical discharges, but it is probably not possible to
unambiguously detect an air shower signal behind the violent radiative background of a thunderstorm. With a combination of
particle detectors and an array of radio antennas electrical activity after an air shower passage could be imaged, allowing for a
study of temporal and spatial coincidences.

\label{sec:con}

\begin{small}  
\vspace*{2ex} \par \noindent
{\em Acknowledgements.}
Part of this research has been supported by Grant number VH-NG-413 of the Helmholtz Association.
\end{small}

\end{document}